\documentclass[a4paper,fleqn,usenatbib]{mnras}

\usepackage{newtxtext,newtxmath}

\usepackage[T1]{fontenc}
\usepackage{ae,aecompl}

\usepackage{graphicx}	
\usepackage{amsmath}	
\usepackage{amssymb}	

\newcommand{\msol}{\text{M}_{\odot}}
\newcommand{\nic}{^{56}\text{Ni}}
\newcommand{\angstrom}{\mbox{\normalfont\AA}}

\title[A 22 $\msol$ Type Ic SN]{Type Ic Supernova of a 22 $\msol$ Progenitor}

\author[Teffs et al.]{
Jacob Teffs,$^{1}$\thanks{j.j.teffs@ljmu.ac.uk}
Thomas Ertl,$^{2}$
Paolo Mazzali,$^{1,2}$
Stephan Hachinger,$^{3}$
Thomas Janka$^{2}$
\\
$^{1}$Astrophysics Research Institute, Liverpool John Moores University, IC2 Liverpool Science Park, 146 Brownlow Hill, Liverpool L3 5RF, UK\\
$^{2}$Max-Planck Institut f\"ur Astrophysik, Karl-Schwarzschild-Str. 1, D-85741 Garching, Germany\\
$^{3}$Leibniz Supercomputing Centre (LRZ), BoltzmannStr. 1, D-85748 Garching, Germany
}

\date{Accepted XXX. Received YYY; in original form ZZZ}

\pubyear{2019}

\begin{document}
\label{firstpage}
\pagerange{\pageref{firstpage}--\pageref{lastpage}}
\maketitle

\begin{abstract}
Type Ic supernovae (SNe\,Ic) are a sub-class of core-collapse supernovae that exhibit no helium or hydrogen lines in their spectra. Their progenitors are thought to be bare carbon-oxygen cores formed during the evolution of massive stars that are stripped of their hydrogen and helium envelopes sometime before collapse. SNe\,Ic present a range of luminosities and spectral properties, from luminous GRB-SNe with broad-lined spectra to less luminous events with narrow-line spectra. Modelling SNe\,Ic reveals a wide range of both kinetic energies, ejecta masses, and  $\nic$ masses. To explore this diversity and how it comes about, light curves and spectra are computed from the ejecta following the explosion of an initially 22 $\msol$ progenitor that was artificially stripped of its hydrogen and helium shells, producing a bare CO core of $\sim$ 5 $\msol$, resulting in an ejected mass of $\sim$ 4 $\msol$, which is an average value for SNe\,Ic.  Four different explosion energies are used that cover a range of observed SNe. Finally, $\nic$ and other elements are artificially mixed in the ejecta using two approximations to determine how element distribution affects light curves and spectra. The combination of different explosion energy and degree of mixing produces spectra that roughly replicate the distribution of near-peak spectroscopic features of SNe\,Ic.  High explosion energies combined with extensive mixing can produce red, broad-lined spectra, while minimal mixing and a lower explosion energy produce bluer, narrow-lined spectra.
\end{abstract}




\section{Introduction} \label{sec:intro}

Stripped envelope supernovae (SESNe) are the explosion following core collapse of stars which experience significant mass loss during their evolution \citep{Woosley1997}. This mass loss may be due to binary interaction or may occur through periods of high mass loss, with rates of 10$^{-4}$ to 10$^{-5}$ $\msol$ yr$^{-1}$ \citep{NOMOTO1995173,Tramper_2016}. If the hydrogen envelope is mostly stripped away the SN is of Type IIb; if the entire hydrogen envelope and parts of the helium are stripped the SN is of Type Ib, and if all of the hydrogen and most of the helium are stripped a Type Ic results \citep{1995ApJ...450L..11F}. With no hydrogen envelope recombining to power a plateau-like phase typical of Type IIP \citep{1994ApJ...430..300E}, the primary source of luminosity in these SESNe is thought to be the radioactive decay of $\nic$. 

\citet{10.1093/mnras/stx980} described a way to sub-classify SESNe starting from two basic classes, He-rich and He-poor SNe.  For He-rich SNe, like Type IIb/Ib, the classification scheme is based on the strength, velocity, and absorption/emission ratio of the H$\alpha$ lines. For He-poor SNe, like Type Ic/Ic-BL, the focus is instead on the number of commonly visible absorption lines, defined as N, present in the spectra at or near maximum luminosity, in the commonly observed wavelength range 4000-8000\,\AA. The number of absorption features in the spectra is dependent on the density and velocity profiles of the line-forming region, the composition of the ejecta, and the epoch of observation. These factors determine the strength and degree of blending of the various features, which can change over the evolution of the ejecta as the line-forming region recedes deeper into the ejecta.  A set of absorption features commonly observed in the spectra of Type Ic SNe, including the three strongest Fe\,II multiplet 48 lines near 5000\,$\angstrom$, Na\,ID\,5895, Si\,II\,6355, the O\,I\,7774 feature  and the Ca\,II NIR triplet are considered \citep{1997ARA&A..35..309F}. These lines are in the optical part of the spectra and are observable from ground-based telescopes, but good S/N is needed to prevent erroneous classifications, especially in the Fe\,II lines.  

A number of methods can be used to determine the basic properties of the explosion, such as ejected mass ($M_{\text{ej}}$), kinetic energy ($E_{\text{k}}$), and the mass of newly synthesized $\nic$, and therefore infer the mass of the progenitor star. The methods described in \citet{1982ApJ...253..785A} (Arnett's Rule) use the light curve of the SNe to estimate the kinetic energy of the explosion and approximate the mass of the ejecta. The light curves of Type Ic SNe can be reproduced with multiple combinations of $E_{\text{k}}$, mass $M_{\text{ej}}$, and the opacity of the ejecta \citep{1998Natur.395..672I}.  Photospheric and nebular spectra of the SNe can be used to refine the $M_{\text{ej}}$ measurement as well as determine the elemental abundances in the ejecta \citep{2005MNRAS.360.1231S,2008MNRAS.386.1897M}. Photospheric spectra are obtained early in the evolution of the SN and thus can only probe the outer layers of the ejecta. The transition from the photospheric phase to the nebular phase occurs approximately 6-12 months after the explosion. At this point the ejecta has expanded and become sufficiently optically thin, so that the spectrum changes from a blackbody-like continuum with absorption features to an emission line spectrum \citep{10.1111/j.1365-2966.2010.17133.x}. Nebular-epoch spectroscopy probes the innermost region of the ejecta. Modelling the spectra at both these stages one can determine the density/abundance structure in the ejecta. This density structure can be used to calculate a mass and combined with the abundance stratification, a synthetic light curve can be computed that can be compared to the observed light curve. 

Evidence from photospheric and nebular spectra shows that the explosion mechanism the ejecta of SESNe are likely asymmetric in nature \citep{Mazzali1284,2008ARA&A..46..433W, Maeda1220,2009MNRAS.397..677T,Tanaka_2009,2011ApJ...739...41C,2015MNRAS.453.4467M,2017MNRAS.469.1897S}. Asymmetry in the explosion mechanism can lead to significant amounts of turbulent mixing, which is probably a key factor in the explosion mechanism of core-collapse SNe \citep{1995ApJ...450..830B,HERANT1995117,1996A&A...306..167J}. Other work regarding the affect mixing has on the light curve and spectra of Type Ib/c has been done by \citet{2012MNRAS.424.2139D} and \citet{2019ApJ...872..174Y}, but focus mostly on the extremely early phases of the SNe.  The extent of this mixing is important in matching models to observations, such as the case of SN\,1987A, which contained high velocity $\nic$ bullets in the hydrogen rich outer ejecta \citep{1988PASAu...7..446H,1990ApJ...360..242S}.  \citet{2019A&A...621A..71T} also require high mixing of $\nic$ to replicate the light curves of broad lined Type Ic SNe. However, mixing this material in a 1-D code is an approximation to an inherently 3-D process and requires a method to emulate the range of possible mixing processes. 3-D simulations can match the observed $\nic$ velocity in SN 1987A  \citep{2018arXiv181211083U} but the time and computing costs of these simulations limit their modelling efficiency.  

In this work we consider an evolved stellar model and explore its predictive ability as a progenitor.  The model is evolved in a single star calculation and prior to collapse it is artificially stripped of its H/He layers, such that the resulting spectra should resemble those of SNe\,Ic. The methodology of the synthetic spectroscopy and photometry is discussed in Section \ref{sec:meth}. We consider the original model, a partially mixed model, and a fully mixed model. In Section \ref{sec:results} we present the bolometric light curves and synthetic spectra generated for all explosion energies and mixing approximations. In Section \ref{sec:phys_class} we classify our synthetic spectra based on the physical classification system previously discussed. In Section \ref{sec:fits}, we compare the models to the time dependent spectra and photometry of several well sampled SNe. In Section \ref{sec:disc} we discuss the results and in Section \ref{sec:conc}, we summarize our results and consider future extensions.

\section{Methods} \label{sec:meth}

In the following sections we describe the methods used to generate the bolometric light curves and synthetic spectra.

\subsection{Progenitor Models} \label{subsec:models} 

The 22 $\msol$ progenitor model used in this work is a modified version of the 22 $\msol$ non-rotating solar metallicity model generated by \citet{RevModPhys.74.1015}. This model is artificially stripped of its outer hydrogen and helium shells as the physical mechanism to remove the mass is not the focus of this work. This leaves a carbon-oxygen (CO) core of $\approx$ 4.5 to 5.25 $\msol$. The mass of the CO core changes as the explosion energy increase due to the explosion mechanism used and is discussed in detail later. After a compact remnant is formed and removed from the model, the final ejected mass is $\approx$ 3.3-4 $\msol$, depending on the explosion energy, as shown in Table \ref{tab:model_params}.  $\alpha$-rich freeze-out in the innermost core produces helium, but no He is present in the atmosphere as any helium in the outermost region that was produced or mixed during the evolution of the star is stripped from the model prior to core collapse.

\begin{table*} 
    \begin{tabular}{ccccc}
    \hline
Energy [$10^{51}$ erg] & $M_{\text{tot}}$/$\msol$ & $M_{\text{ej}}$/$\msol$ & $M_{\text{Ni}}$/$\msol$	& $E_{\text{k}}$/$M_{\text{ej}}$ (10$^{51}$ ergs / $\msol$) \\
\hline
\hline
1	   & 4.46 & 3.35 & 0.097 & 0.29 \\
3	   & 5.15 & 3.74 & 0.147 & 0.80 \\
5	   & 5.26& 4.01 & 0.187 & 1.24 \\
8      & 5.3 & 4.05 & 0.224 & 1.97 \\
    \hline
\end{tabular}
\caption{The kinetic energy, total mass, ejecta mass, $\nic$ mass, and the ratio $E_{\text{k}}$/$M_{\text{ej}}$ for the four models used in this work. The difference in total and ejected mass is related to the explosion mechanism used and is discussed in Section \ref{subsec:models} along with the possibly uncertain $\nic$ mass. }
\label{tab:model_params}
\end{table*}

Some SNe\,Ic, such as SN\,1998bw, may eject as much as 8-10 $\msol$ of material with $E_{\text{k}}$ of several $10^{52}$\,erg after the stripping of a 30-50 $\msol$ progenitor \citep{1998Natur.395..672I,Galama1998,Mazzali_2001}, while others, such as SN\,1994I, have ejecta masses estimated to be $\sim 1-2$ $\msol$ and $E_{\text{k}} \sim 1-2$ $\times$ $10^{51}$\,erg \citep{2006MNRAS.369.1939S}. \citet{10.1093/mnras/sty3399} estimated ejecta masses using Arnett's rule and found that SNe\,Ib/c have a broad range of masses. Using this method, GRB-associated SNe\,Ic have a mean $M_{\text{ej}} = 4.7 \pm 1.5$ $\msol$, while more narrow-lined SNe\,Ic have a mean $M_{\text{ej}} = 3.2 \pm 2.4$ $\msol$. The mass ejected in our models is thus a ``mean" SN\,Ic mass, but it does not represent any group in particular, and is only the first step in our effort to model various progenitor masses. 

Using the physical classification system from \citet{10.1093/mnras/stx980}, GRB-SNe have $N=3-4$, while more narrow-lined SNe have $N=6-7$, where $N$ is defined as the number of visible features, such that increased blending leads to a smaller value of $N$. The value of $N$ is set primarily by $E_{\text{k}}$. Just like $M_{\text{ej}}$, $E_{\text{k}}$ also covers a wide range. In order to represent this spread of $E_{\text{k}}$, we exploded our 22 $\msol$ progenitor with four different energies. 

\begin{figure}
\includegraphics[scale=.83]{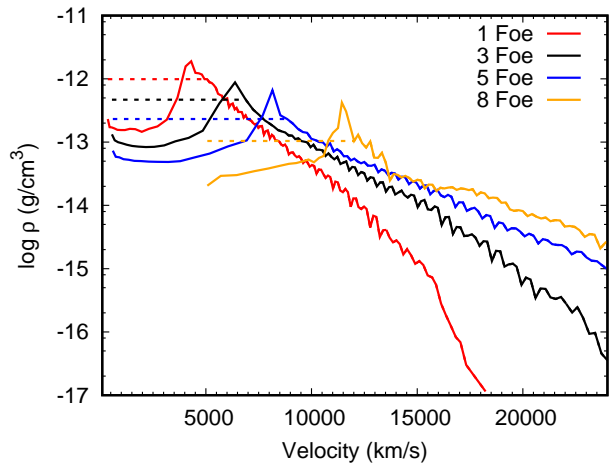}
\caption{The density profiles of the four explosion models at the onset of homologous expansion, or approximately 5 days after the onset of the explosion (fully drawn lines). The dashed lines show the flattened density structures used for the fully/partially mixed models. The mass of the flattened region is 0.997, 1.33, 1.35, and 1.66 $\msol$ for the 1, 3, 5, and 8 foe models, respectively. The majority of this mass is contained with the density peaks visible in the original density structures. The jagged behaviour of the density profiles is magnified by the logarithmic scaling and does not affect the evolution as much as the primary density spike. }
       \label{fig:dens}
\end{figure}

The supernova explosion is simulated with a 1-D hydrodynamics code with neutrino transport called \textsc{Prometheus-HOTB}, described in more detail in \citet{1996A&A...306..167J} and \citet{0004-637X-757-1-69}. The neutrino heating used to generate the explosion is artificially enhanced in order to achieve final explosion energies of 1, 3, 5, and $8 \times 10^{51}$ ergs, respectively. This energy is deposited after core bounce in the surroundings of the newly formed neutron star over a timescale of several seconds. This is supposed to mimic, in a parametric way, the dynamical consequences of an engine that generates thermal energy by neutrinos or magnetic field effects as required to explode the 22\,$\msol$ star at the chosen $E_{\text{k}}$.  The density profiles of the four models are shown in Fig. \ref{fig:dens}. They are characterized by distinctive spikes, which mark the outer boundary of the region where explosive nucleosynthesis affects the composition.

The density inversion is a consequence of the fast neutrino-driven wind, which is stronger for more energetic explosions. It pushes the slower ejecta expanding behind the outgoing supernova shock and accumulates them in a dense shell surrounding a lower-density central bubble. This central bubble is filled by high-entropy wind matter and can be recognized in Fig. \ref{fig:dens}, at a much later stage when the ejecta have reached homologous expansion, by the density trough around the coordinate origin. In 1D simulations including nickel decay the density spike is inflated by the radioactive decay heating \citep{2018MNRAS.475..277J}. In 3D simulations it is further smoothened by Rayleigh-Taylor mixing of the ejecta, see Figure 8 in \citet{2017ApJ...846...37U}; compare panels b and c there at times $t < 1\,$hour, i.e. before the reverse shock from the He/H interfaces moves inward through the expanding stellar debris.

This spike in the density profile results in a relatively large amount of material in a small velocity range. For the 1 foe model, the density spike ranges from approximately 3000 to 5500 km~s$^{-1}$ and contains 1.6 $\msol$ of material while the 8 foe model has a similar $\approx$ 3000 km~s$^{-1}$ range starting from 10500 to 13600 km~s$^{-1}$ that contains 1.24 $\msol$ of material. 

The model is evolved in \textsc{Prometheus-HOTB} until the shock leaves the star and the ejecta are in homologous expansion, such that $r \propto v t$. As part of the explosion mechanism, the final ejected masses in the various models differ somewhat (see Table \ref{tab:model_params}). This is because a larger mass of neutrino-heated ejecta is needed to account for higher values of $E_{\text{k}}$.  
\begin{table*} 
    \begin{tabular}{cccccc}
    \hline
Energy [$10^{51}$ erg] & M$_{\text{He}}$/$\msol$ & $M_{\text{Ni}}$/$\msol$ & M$_{X_\text{56}}$/$\msol$	& 50\% M$_{X_\text{56}}$/$\msol$ \\
\hline
\hline
1	   & 0.059 & 0.083 & 0.029 & 0.0147 \\
3	   & 0.119 & 0.090 & 0.114 & 0.057 \\
5	   & 0.180 & 0.079 & 0.215  & 0.107 \\
8      & 0.245 & 0.066 & 0.315  & 0.158 \\
    \hline
\end{tabular}
\caption{The total mass of He, $\nic$, the X$_{\text{56}}$ tracer used to track the iron-group rich material, and 50\% of X$_{\text{56}}$ that is added to the $\nic$ mass. }
\label{tab:xfe_ratio_issues}
\end{table*}

For explosive nucleosynthesis, the code uses a 13-species $\alpha$-network and a 15-species solver for nuclear statistical equilibrium that tracks the bulk production of $\alpha$-elements up to $\nic$. $\nic$ production is exclusively confined to the innermost region.  The $\nic$ masses listed in Table \ref{tab:model_params} include the $\nic$ that is produced explosively by nuclear burning in shock-heated ejecta as well as 50\% of the so-called tracer material. This material, defined as X$_{56}$, accounts for iron-group nuclei in slightly neutron-rich neutrino-heated ejecta, whose exact composition depends on details of the neutrino transport. Its exact composition is therefore uncertain because of the use of an approximate description of the neutrino physics in the explosion simulations \citep{Ertl_2016,2016ApJ...821...38S}. Therefore, we take X$_{\text{Ni}}$ + 50\% of X$_{56}$ to be a reasonable approximation for the $\nic$ mass.

Table \ref{tab:xfe_ratio_issues} shows the mass of the He, $\nic$, and the X$_{\text{56}}$. The explicitly tracked $\nic$ formation is under 0.1 $\msol$ for all four explosion energies while the  X$_{56}$ increases by almost 0.1 $\msol$ for each 'jump' in $E_{\text{k}}$. The total $\nic$ mass used for the light curve and spectral calculations (column 4 in Table \ref{tab:model_params}) is increasingly comprised of the X$_{56}$ tracer instead of $M_{\text{Ni}}$ as $E_{\text{k}}$ increases, as shown in columns 3 and 5 of Table \ref{tab:xfe_ratio_issues}.


\subsection{Abundance Profiles} \label{subsec:abu_prof}

\begin{figure}
\includegraphics[scale=.8]{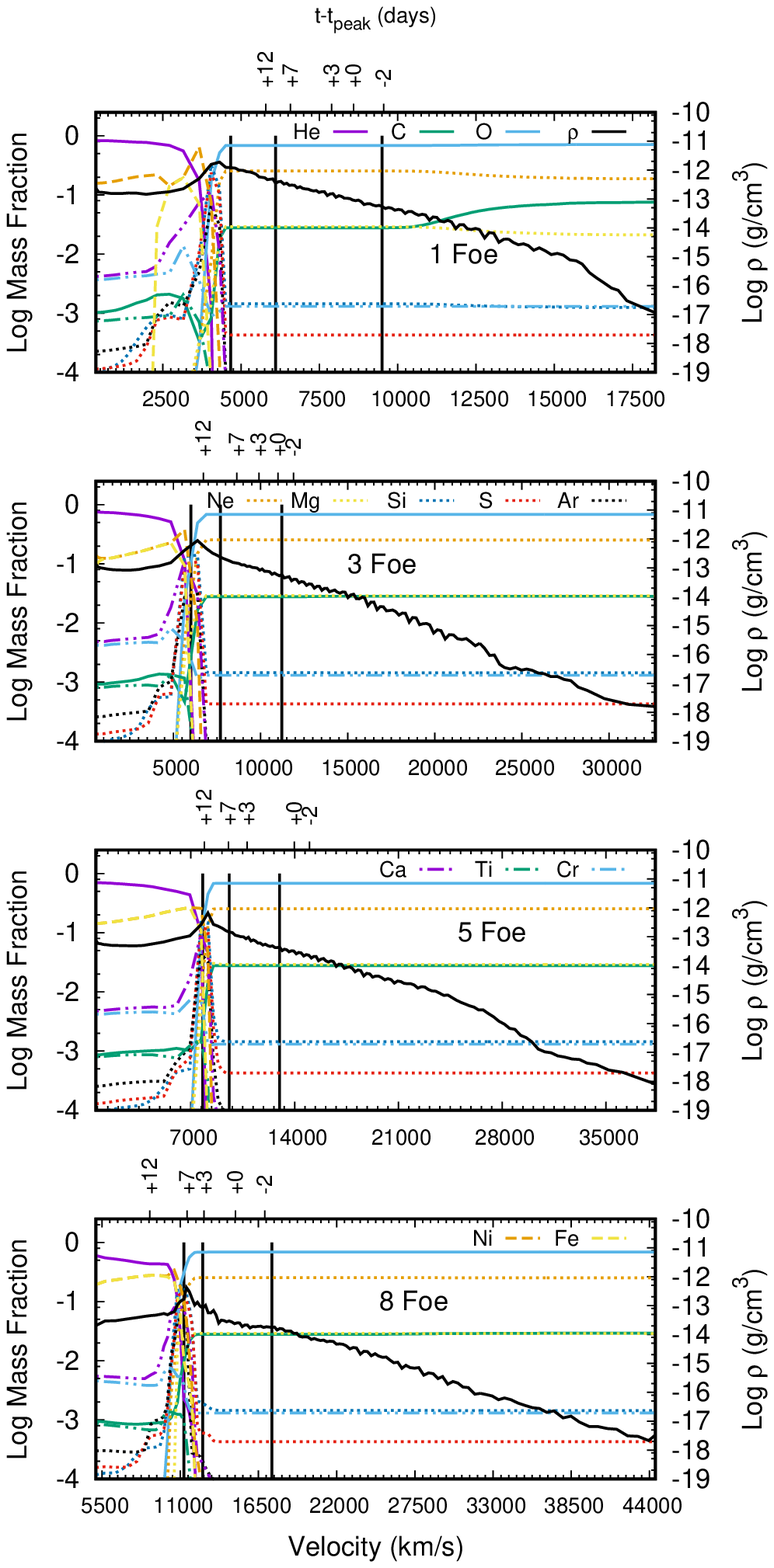}
\caption{The abundance (left-hand scale) and density (right-hand scale) profiles of the 1, 3, 5, and 8 foe models (from top to bottom, respectively) with no mixing applied at approximately 5 days after the onset of the explosion. The Fe line represents Fe and the other 50 $\%$ of the X$_{56}$ tracer not included in the $\nic$ abundance.  Notice that the x-axes of the four plots have different ranges to account for the different $E_{\text{k}}$. Tick marks at the top of each panel represent the position of the photosphere at the epochs indicated (relative to bolometric maximum). The black vertical lines define the velocities within which 1, 2, and 3\,$\msol$ of material are enclosed (from low to high velocities). }
\label{fig:abu_all}
\end{figure}

\begin{figure}
\includegraphics[scale=.8]{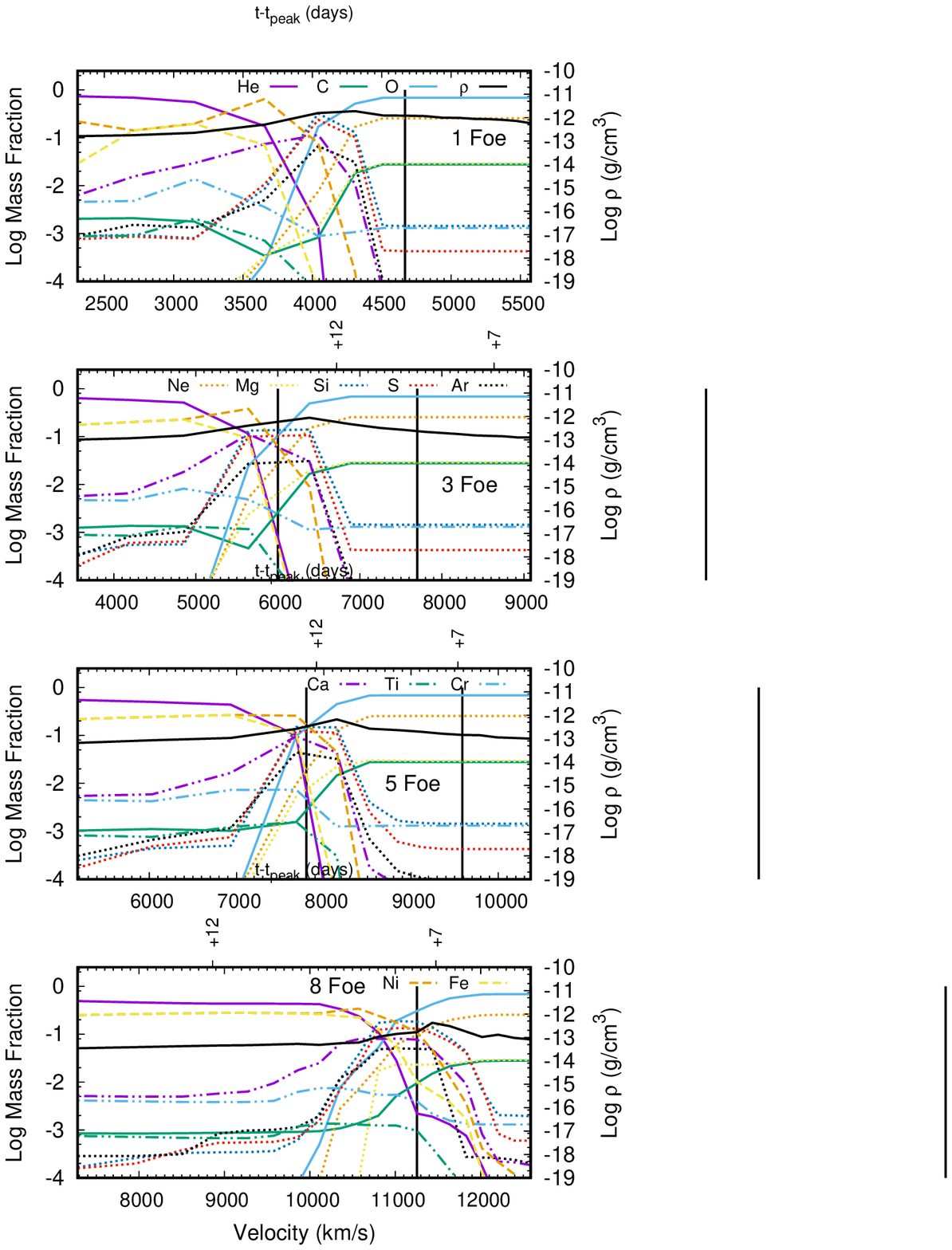}
\caption{The inner region of the non-mixed models at approximately 5 days after the onset of the explosion are shown here as the velocity range is too narrow in Figure \ref{fig:abu_all}.  Symbols and labels are the same as in Fig. \ref{fig:abu_all}.}
    \label{fig:abu_all_inner}
\end{figure}

The abundance distribution of the original models is shown in the four panels of Fig. \ref{fig:abu_all}. An expanded view of the regions in and near the density spikes is shown in Fig. \ref{fig:abu_all_inner}. In the original models, $\nic$ is confined to the lowest velocities, which is known not to be realistic. In most SNe\,Ic that have been modelled, it is necessary for $\nic$ to be present at intermediate/high velocities in order for synthetic light curves to match the observed ones.  The outer part of the ejecta is comprised predominantly of C, Ne, O, and Mg, with the bulk of the intermediate mass elements (IME) produced at the edge of the $\nic$-forming region. $\nic$ produced by the explosion is contained within and behind the density peak near 4000, 6000, 9000 and 11000 km~s$^{-1}$ for the four models, in order of increasing $E_{\text{k}}$.  

\begin{figure}
\includegraphics[scale=.8]{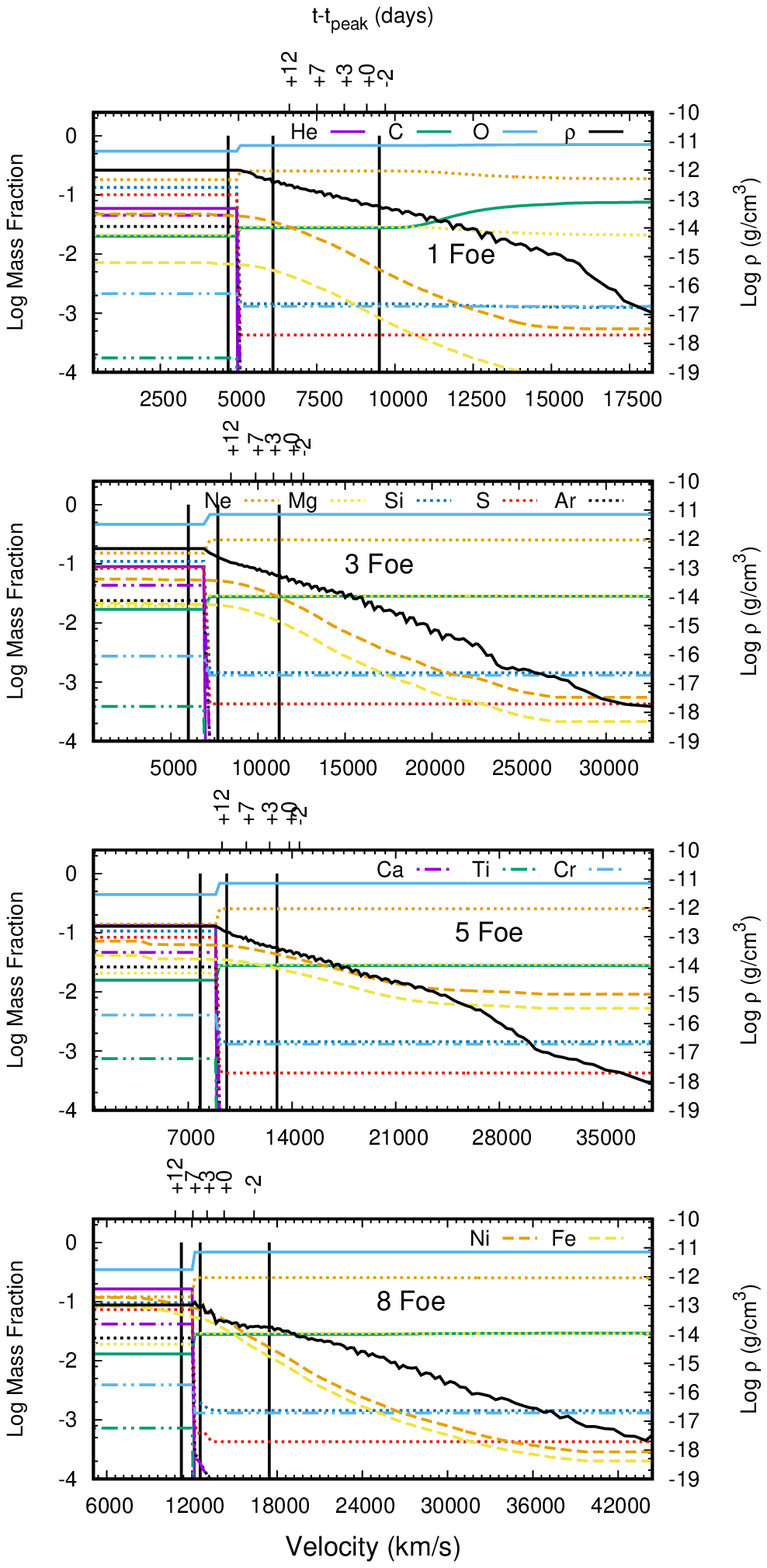}
\caption{The abundance (left-hand scale) and density (right-hand scale) profiles of the 1, 3, 5, and 8 foe partially mixed models (from top to bottom, respectively) at approximately 5 days after the onset of the explosion. The Fe line represents Fe and the other 50$\%$ of the X$_{56}$ tracer not included in the $\nic$ abundance. Notice that the x-axes of the four plots have different ranges to account for the different $E_{\text{k}}$. Tick marks at the top of each panel represent the position of the photosphere at the epochs indicated (relative to bolometric maximum). The black vertical lines define the velocities within which 1, 2, and 3\,$\msol$ of material are enclosed (from low to high velocities). }
\label{fig:abu_mixing} 
\end{figure}

\begin{figure}
\includegraphics[scale=.8]{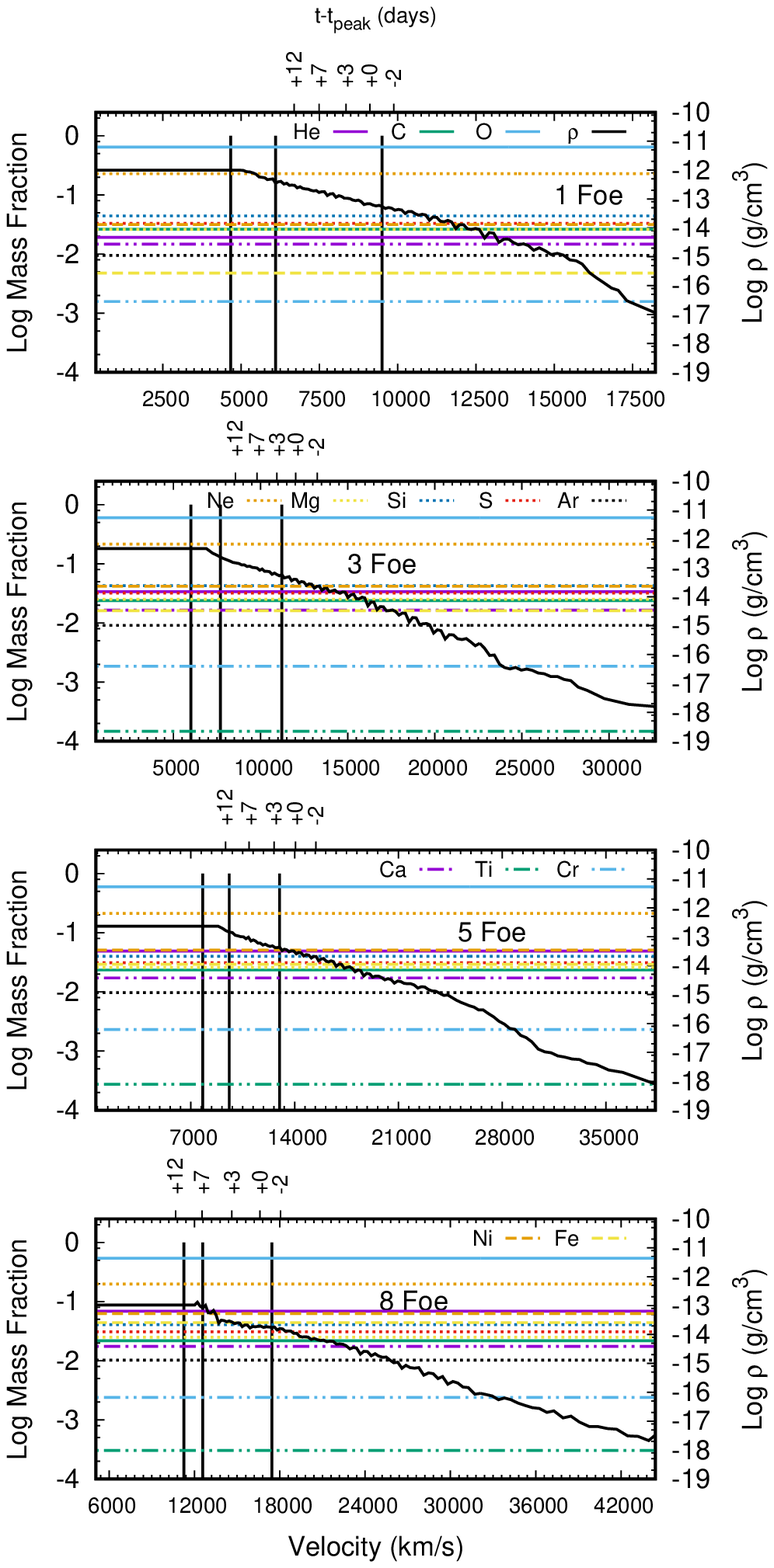}
\caption{The abundance (left-hand scale) and density (right-hand scale) profiles of the 1, 3, 5, and 8 foe fully mixed models (from top to bottom, respectively) at approximately 5 days after the onset of the explosion.  The Fe line represents Fe and the other 50$\%$ of the X$_{56}$ tracer not included in the $\nic$ abundance. Notice that the x-axes of the four plots have different ranges to account for the different $E_{\text{k}}$. Tick marks at the top of each panel represent the position of the photosphere at the epochs indicated (relative to bolometric maximum). The black vertical lines define the velocities within which 1, 2, and 3\,$\msol$ of material are enclosed (from low to high velocities).}
\label{fig:abu_full}
\end{figure}

In order to explore the properties of element mixing, we use three approximations in this work. First we assume that the original model explodes ``as is" and nothing is changed. We call these models ``unmixed" or ``non-mixed" with the abundance profiles of these models shown in Figures \ref{fig:abu_all} and \ref{fig:abu_all_inner}. A second approximation is to mix the region that undergoes explosive nucleosynthesis in both density and abundances. The range of velocities that this region encompasses changes with $E_{\text{k}}$. To account for this, we choose a boundary point just beyond the peak of the density structure, shown in Figure \ref{fig:dens}, that defines our density and compositional mixing region. The result is a flat density structure and constant composition. Similar to the original model, this mixing approach would keep $\nic$ centrally located, which we know is unlikely to be correct. To mix out this material, a running boxcar average is used to only the mix the $\nic$ and Fe abundances in ejecta. This results in a partially centrally located $\nic$ with decreasing $\nic$ further into the outermost regions. We call these models ``partially mixed" and the abundance profiles are shown in Figure \ref{fig:abu_mixing}. In a third approximation we assume that the ejecta are completely mixed, while keeping the density structure as in the second approximation. We call these models ``fully mixed" and their abundance profiles are shown in Figure \ref{fig:abu_full}.

The unmixed models contain some helium at the lowest velocities (Fig. \ref{fig:abu_all} and \ref{fig:abu_all_inner}). This is produced during $\alpha$-rich freeze-out. Complete mixing extends this material to the entire range of velocities, as seen in Fig. \ref{fig:abu_full}.  While it is hard to imagine that such extreme mixing is feasible for a 1 foe explosion of a 3.35 $\msol$ CO core, we treat it as a theoretical endpoint in the mixing approximations. Extremely mixed ejecta may be linked with highly asymmetric explosions and possibly related to gamma-ray bursts or hypernovae, which often have larger ejecta masses and explosion energies \citep{2004ApJ...614..858M,10.1093/mnras/stz1588}.

Table \ref{tab:xfe_ratio_issues} shows the total mass of He for each $E_{\text{k}}$. Because of the choice of outer mass-cut, no He is located in the outer layers of the ejecta (Fig. \ref{fig:abu_all}). The partially mixed models do not mix beyond the innermost region and the He abundance outside th
e mixing region is poor or non existent. Only in the fully mixed models does helium extend to the outer layers and higher velocities.  For the fully mixed models, the mass fraction of the He is $\sim 10^{-3}$. For the 1, 3, and 5 foe fully mixed models, the latest spectra are at $t - t_{\text{peak}}$ +12 days and is either at or above the 2 $\msol$ line in the models. This leaves approximately 1.5 - 2 $\msol$ of material in the line-forming region of which 0.02 - 0.04 $\msol$ of this is He. \citet{2012MNRAS.422...70H} showed that approximately 0.06 - 0.14 $\msol$ of helium in the line-forming region is required to form He lines in the spectra. Given that He line formation is sensitive to the location of $\nic$, this suggests that despite the low mass of He present in the line-forming region, the extra $\nic$ may lead to the formation of He lines via non-thermal processes \citep{1991ApJ...383..308L,2012MNRAS.422...70H}. However, the extensive mixing also places $\nic$ alongside this helium. As such, the fully mixed models will include the non thermal affects of He regardless of the approximate mass in the line forming regions.

For the 8 foe unmixed model, the latest epoch places the photosphere deep in the innermost region where the abundance of He is $\sim 40$\%, shown in Figure \ref{fig:abu_all}. The mass of the innermost core is $\sim 1$ $\msol$, but only 25-50\% of this mass is above the photosphere with the abundance of He dropping towards the density spike. This puts approximately 0.1 - 0.2 $\msol$ of He in the line forming region of the latest epoch of non mixed 8 foe model. This layer is in a region in which the elemental abundances are comprised of approximately 40$\%$ He, 30$\%$ Fe, and 30$\%$ Ni, which may not be a realistic or observable layer.

\subsection{Light Curve and Spectral Synthesis Codes} 
\label{subsec:lc_specsyn}

After the explosions have entered the homologous expansion phase and we apply one of the three mixing approximations, we then continue the simulation workflow with our light curve code. The code, described in detail in \citet{1997A&A...328..203C}, calculates the emission and propagation of gamma rays and positrons produced by the decay of $\nic$ and, subsequently, $^{56}$Co, into the homologously expanding ejecta using a Monte Carlo method. Constant values for both positron and gamma-ray opacity are used, with values of 7\,cm$^2$g$^{-1}$ and  0.027\,cm$^2$g$^{-1}$, respectively \citep{1980PhDT.........1A}. The code does not take into account post-breakout emission phase. The energy that is deposited is then recycled into optical photons, whose propagation is also followed in a Monte Carlo scheme.  A time- and metallicity-dependent but frequency independent optical opacity is used \citep{2001ApJ...547..988M}. This aims at reproducing the dominance of line opacity in the ejecta \citep{10.1093/mnras/stz1588}. A constant time step of 1 day is used and the calculation is run to 200 days. The position of the photosphere is approximately determined by integrating inwards until the radius where an optical depth of $\tau \ge 1$ is found. The zone that includes this radius is used to determine the photospheric velocity. This velocity, combined with the luminosity, abundances, and epoch are used as inputs in the spectral code. 

The synthetic spectra are calculated using a Monte Carlo code discussed in detail in \citet{1993A&A...279..447M, 1999A&A...345..211L, 2000A&A...363..705M}. The code reads in the density structure as a function of velocity, stratified composition, luminosity, and velocity of the photosphere at a given epoch. When modelling observed SNe, these variables can be inferred by fitting the observed spectra and light curves. As the photosphere recedes into the expanding ejecta, the near-photospheric region, where line formation is most likely to occur, can be characterized by a varying composition.  A detailed depth-dependent composition can be inferred by matching the spectra in a process called abundance tomography \citep{2005MNRAS.360.1231S}. If the ejecta has He or H in abundances greater than 10$^{-10}$, the non-thermal effects of these elements are included into the code by the use of a non-local thermodynamic equilibrium (NLTE) module discussed in \citet{2012MNRAS.422...70H}. The formation of H/He lines results from significant deviations from LTE as shown in detail by \citet{1991ApJ...383..308L}. For these SN\,Ic models, the progenitor was stripped such that the outer layers contain an abundance of helium smaller than 10$^{-10}$, but the $\alpha$-rich freeze-out results in some He mass as discussed in Section \ref{subsec:abu_prof}

For all models, $t_{\text{peak}}$ is defined as the time when the bolometric light curve reaches a maximum luminosity ($L_{\text{max}}$). For easier comparison among the models, synthetic spectra for all energies and mixing approximations are generated at the same five epochs relative to $t_{\text{peak}}$ rather than at fixed epochs relative to the time of explosion. This is because  $t_{\text{peak}}$ occurs at different times after explosion for different models owing to the different $E_{\text{k}}$, mass and distribution of $\nic$. We choose times of -2, 0, +3, +7, and +12 days with respect to $t_{\text{peak}}$. These times are chosen to cover $\sim 2$ weeks of observations, in order to capture the evolution of the main spectral features and match the epochs of most available data. Observationally, well sampled SN spectra prior to peak are rarer than post-peak so more spectra are generated after $t_{\text{peak}}$ \citep{2019MNRAS.482.1545S}. Extensions to this would be considering nebular spectra generated at 180 - 220 days post-peak.

\section{Results} 
\label{sec:results}

\subsection{Light Curves} 
\label{subsec:lcs}

\begin{figure}
\includegraphics[scale=.33,angle=-90]{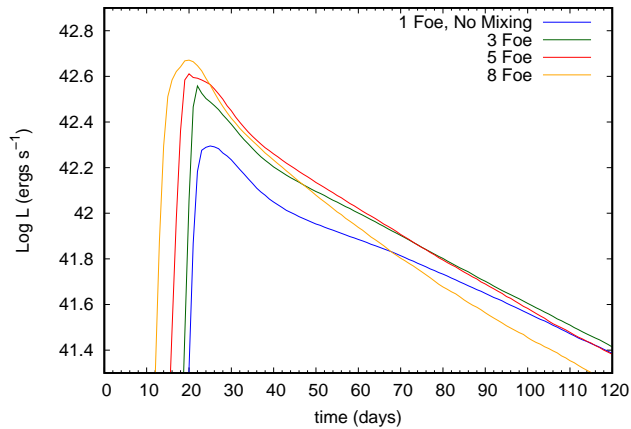}
\caption{Synthetic light curves for the unmixed models for all four energies. 
\label{fig:lc_nomix}}
\end{figure}

\begin{figure}
\includegraphics[scale=.33,angle=-90]{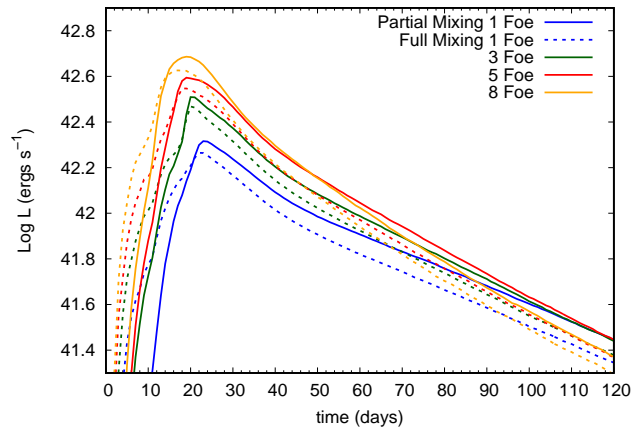}
\caption{Synthetic light curves for the mixed models light curves for all four energies. In contrast to the unmixed models, these models have much broader light curves and longer t$_{1/2}$.
\label{fig:lc_mix}}
\end{figure}

Figures \ref{fig:lc_nomix} and \ref{fig:lc_mix} show the bolometric light curves generated for the three mixing approximations for each of the four energies. The density spike separating the $\nic$ production and the rest of the CO core, shown in Figures \ref{fig:dens} and \ref{fig:abu_all}, acts as a barrier to photon diffusion and causes the rise time to be at least 18 days for all unmixed models. The mixed model tend to have shorter rise times because the density spike is smoothed out and $\nic$ is distributed to higher velocities. Another parameter used to describe SESNe is $t_{\pm 1/2}$, defined as the time it takes for the luminosity to rise from $L_{\text{max}}/2$ to $L_{\text{max}}$ for $t_{-1/2}$ or to decline from $L_{\text{max}}$ to $L_{\text{max}}/2$ for $t_{+1/2}$ \citep{10.1093/mnras/sty3399}. The unmixed models have short $t_{-1/2}$, because initially photons cannot diffuse out because of the density spike, but when  density decreases sufficiently by expansion the release of trapped photons is immediate. Therefore, in the unmixed models $t_{+1/2}$ $>>$ $t_{-1/2}$, while both sets of mixed models are more symmetric about L$_{\text{max}}$. Smoothing the density spike and mixing the $\nic$ outwards allows a much smoother rise time for all energies, resulting in earlier $t_{\text{peak}}$ but a larger $t_{-1/2}$ indicating a slower rise. 

The models in this work do follow Arnett's models to an extent but we do not expect a replication of the exact results. Arnett assumes a point-like distribution of $\nic$, which is not realistic \citep[see, e.g.,][]{10.1093/mnras/stx992} and is not what we use. As $\nic$ is produced further out with increasing $E_{\text{k}}$, diffusion of photons is easier and peak is reached earlier. The declining part, on the other hand, tends to follow Arnett more closely, as the dominant factor that controls opacities then is indeed $E_{\text{k}}$.

\subsection{Spectra} 
\label{subsec:spec}



\begin{figure*}
\centering
\includegraphics[scale=.57,angle=-90]{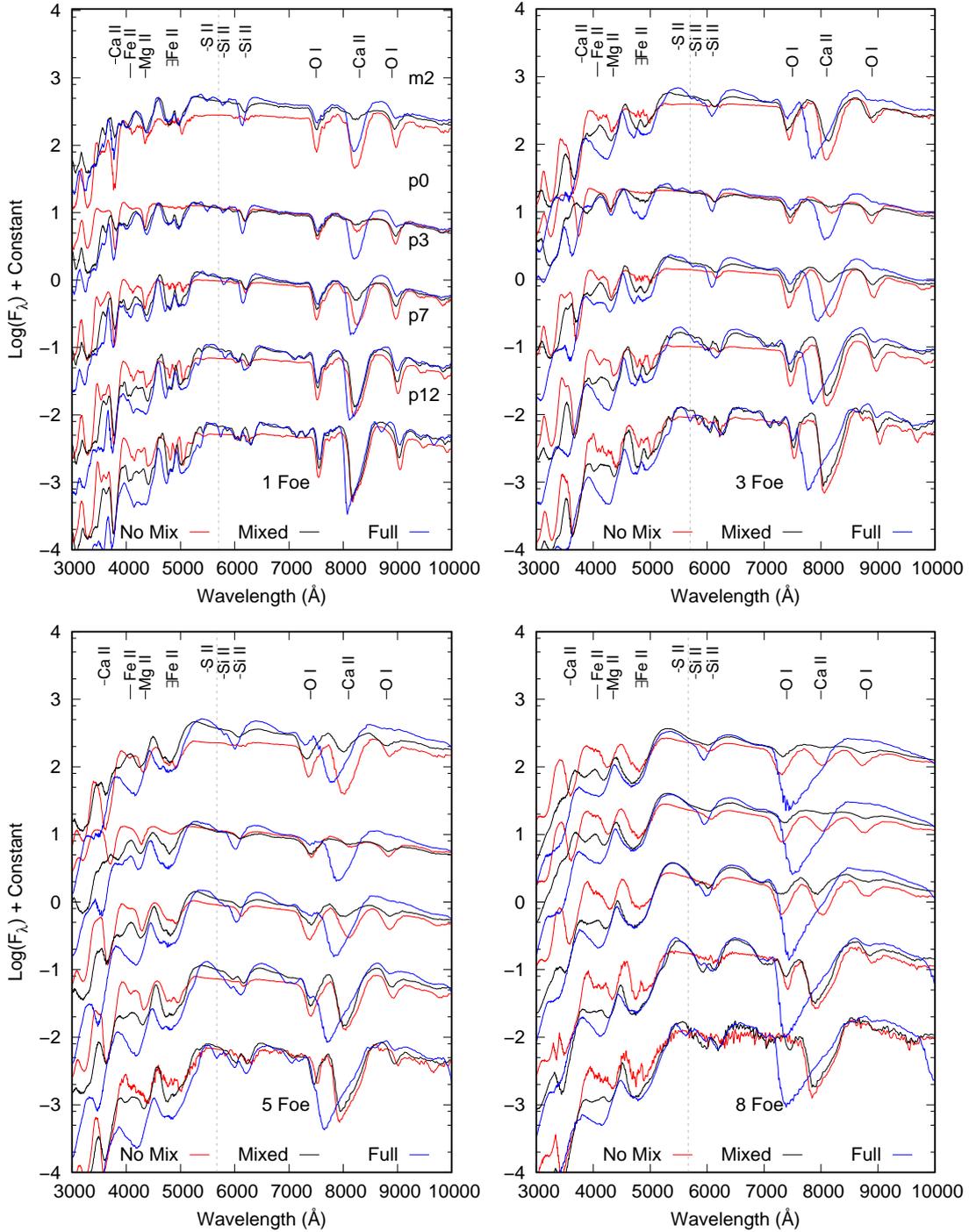}
\caption{The synthetic spectra for all mixing approximations and $E_{\text{k}}$. The flux is in log scale and offset for visibility. The dashed grey line represent the strongest optical He I line at 5875$\angstrom$. All line identifiers are approximate and may not represent the exact minima for the identified spectral feature. \label{fig:spec_all_mix_all_ene}}
\end{figure*}

\begin{figure}
\centering
\includegraphics[scale=.65,angle=0]{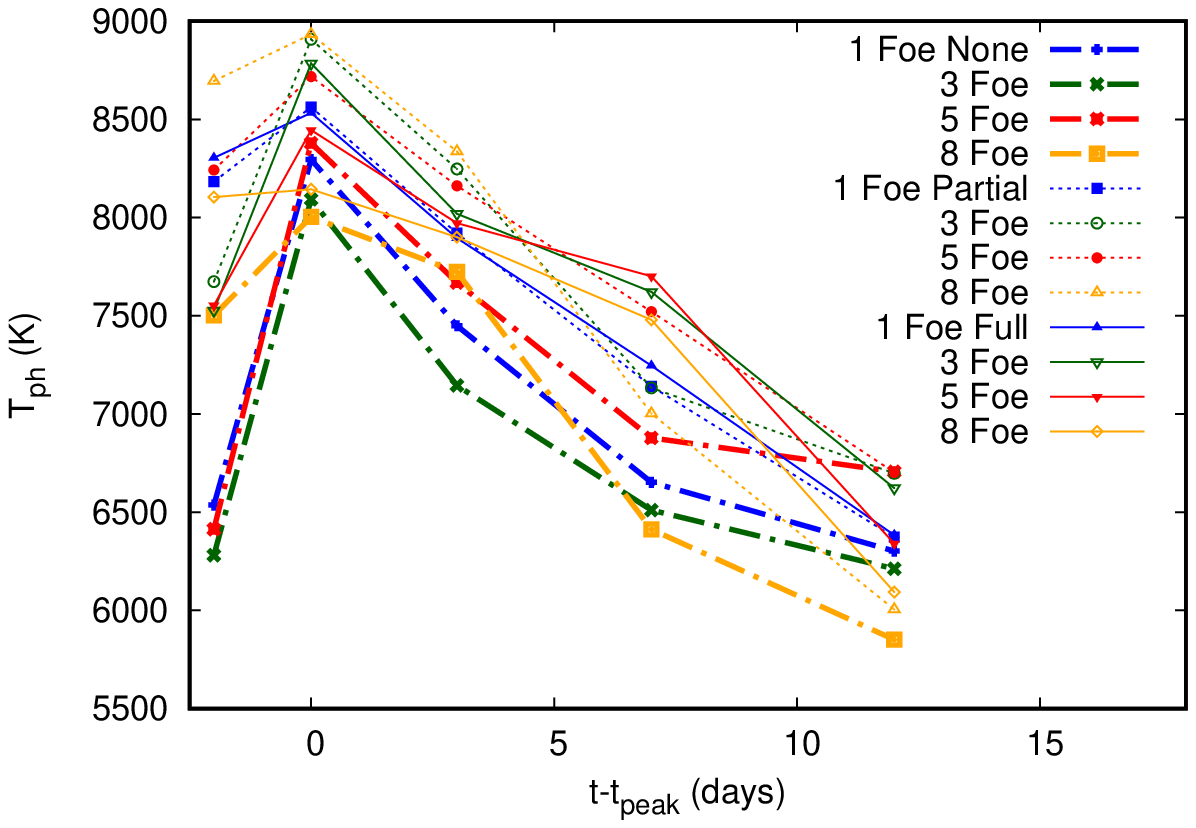}
\caption{The evolution of the photospheric temperatures for all combinations of explosion energy and the mixing approximations.
\label{fig:radtemp}} 
\end{figure}

The spectral evolution of the 1 foe models for all three mixing approximations is shown in the top left plot in Figure \ref{fig:spec_all_mix_all_ene}. The photospheric temperatures at each epoch are shown in Figures \ref{fig:radtemp}. As mixing increases from no mixing to fully mixed, the wavelength range of $\lambda$ < 4500 $\angstrom$ begins to lose features because the increasing abundance of Fe-group elements in the line-forming region causes more line blanketing. This is less noticeable in the early phases, but at $t_{+7}$ and $t_{+12}$ this region flattens noticeably.  The broadness of many optical features also increases as a function of mixing, with several lines becoming indistinguishable. The S\,II and Si\,II lines near 5500-5800\,\AA, on the other hand, are visible only in the fully mixed case. Both S and Si are produced deep in the ejecta and only mixed outwards significantly in the fully mixed case. The Fe\,II multiplet 48 lines near 5000\,$\angstrom$ are split owing to the relatively low velocity of the ejecta, such that all three strongest lines can be identified at some epoch in the evolution of the spectra for all three mixing approximations. The fully mixed model includes the non-thermal effects of He and the 5875$\angstrom$ line is present in the last two epochs with two weaker lines in the t$_{\text{peak}}$ + 12 days at 6500 and 6700 $\angstrom$. Prior to these epochs, the total mass of He in the line forming region is too small to produce strong or easily observable lines. He lines visible in the NIR region of the spectrum are discussed later. 


The spectral evolution of the 3 foe models for all three mixing approximations is shown in the top right plot of Figure \ref{fig:spec_all_mix_all_ene}. These are similar to the 1 foe models with respect to the presence of features. The breadth and visibility of several features increases as the amount of mixing increases, as does line blanketing in the bluer portion of the spectra.  The Fe\,II lines near 5000\,$\angstrom$ are not as easily identifiable in the fully mixed case, as the higher energy leads to higher velocities, producing line blending.  The ejecta do not have enough velocity or energy to blend O\,I 7774 and the Ca\,II IR triplet, even under the full mixing approximation. Similar to the 1 foe fully mixed model, only the day 7 and 12 spectra show evidence of He in the fully mixed model at this $E_{\text{k}}$. Unlike the 1 foe model, the region where the 6500 and 6700 $\angstrom$ He lines would form has two much stronger Fe II lines that make observing the weaker He I lines challenging.


The bottom left plot in Figure \ref{fig:spec_all_mix_all_ene} shows the time-series of spectra for the 5 foe models for all three mixing approximations. Similar to the 1 and 3 foe models, mixing alters the breadth and visibility of several features as well as increasing line blanketing in the near-UV. The Fe\,II lines are now mostly blended together as the ejecta velocities are much higher.  The O\,I 7774 and Ca NIR triplet near 7500 - 8000\,$\angstrom$ are much broader and begin to show signs of blending, at least in the fully mixed case when more calcium is present at high velocity. Continuing the trend, only the day 7 and 12 spectra show He I at 5875$\angstrom$, however the overall blending and broadening of the spectra in general makes the day 7 He I line at 5875$\angstrom$ less discernable from the rest of the spectra.
 

The bottom right plot in Figure \ref{fig:spec_all_mix_all_ene} shows the time-series of spectra for the 8 foe models for all three mixing approximations. The higher ejecta velocity produces consistent blending of the three Fe\,II multiplet 48 lines for all the mixing approximations and epochs chosen. The fully mixed case also blends together the Ca\,II IR triplet and the O\,I 7774 lines due to the higher energy ejecta and the mixing resulting in Ca and O across the entire velocity range. The partially mixed and unmixed models do not mix out the calcium present at the edge of the $\nic$-rich region to the same velocity where oxygen is dominant. The calcium abundance in the outer atmosphere is $\sim 10^{-5}$ in these models, which is too little to cause strong features. The regions where the calcium abundance is significant has velocities that are too low for blending to occur. The near-UV line blanketing is stronger in the fully mixed models as the metal-rich core is mixed into the outer layers. This is particularly strong in comparison to the fully mixed models with lower $E_{\text{k}}$. This shows that $E_{\text{k}}$ $\approx$ 5 $\times$ 10$^{51}$\,erg (or higher depending on mixing) is required for significant blending to occur. This could be seen as the lower boundary for hypernovae. The He I line at 5875$\angstrom$ is only partially observable at day 7 and stronger at day 12 in the fully mixed models. 

At t$_{\text{peak}}$ + 12 days for the 8 foe non-mixed model, the photosphere has receded into the core, which is comprised of 40$\%$ He, 30$\%$ $\nic$ and 30$\%$ Fe. This is shown in the 8 foe models at the bottom of Figures \ref{fig:abu_all} and \ref{fig:abu_all_inner}. At this late epoch, approximately 30 days after the start of the explosion, some of the $\nic$ has decayed into $^{\text{56}}$Co. This material is contained in a very narrow velocity range ($\sim \Delta$v 2500 km~s$^{-1}$) just above the photosphere. The proximity of this region to the photosphere results in the formation of a high number of narrow Fe/Co/Ni lines. Such a narrow-lined spectrum is not observed in Type Ic-BL SNe, of which the 8 foe model would likely be classified, suggesting that the models with a non-mixed Fe/Ni/Co core do not reflect realistic SNe at some or most epochs.

For all values of $E_{\text{k}}$, mixing of Fe group elements and $\nic$ outward produces stronger line blanketing. The Si\,II features near 6000\,$\angstrom$ are consistently weak in the unmixed case as Silicon is located too deep in the ejecta, formed either by the progenitor or during the explosive nucleosynthesis. In the later epochs of the mixed models, the 6000\,$\angstrom$ region near the stronger Si\,II and S\,II lines show the formation of several Fe\,II lines. These are separable and identifiable in the lower energy models at the late epochs due to the ejecta have insufficient energy to blend these lines.    


The Ca abundance at $t_{-2}$ days for the unmixed and partially mixed models is approximately $\sim 10^{-5}$, below the lower limits of the abundance plots in Section \ref{subsec:abu_prof}, while the abundance of Ca in the fully mixed is $\sim 10^{-2}$. In the fully mixed Ca produces a strong and broad Ca\,II NIR triplet line at all epochs. The ionization ratios of Ca in the non-mixed and partially mixed models are different, resulting in different strengths of this line at the $t_{-2}$ epoch. In the unmixed models 1-10\% of Ca is Ca\,II, while in the partially mixed models only 0.1-1.0\% of Ca is Ca\,II. Therefore, the unmixed show in a stronger Ca\,II line. Ionization  ratios are related to both the density and luminosity variations for each model and change as a function of time.


Mixing the $\nic$ and Fe-group elements further out into the ejecta is noticeable in the spectra. The additional Fe-group elements form lines in the bluer part of the spectra, blanketing the other lines that are visible in the un-mixed case as the photosphere recedes deeper into the ejecta near t$_{\text{peak}}+12$ days. The Ca II H \& K lines near 4000\,$\angstrom$ are an example of this behaviour as the bluer side of these lines is blended at higher explosion energies with increased mixing. This line blanketing effect is also correlated with explosion energy as the spectrum blue of 4000 $\angstrom$ is nearly devoid of identifiable lines.

\begin{figure}
\includegraphics[scale=.65]{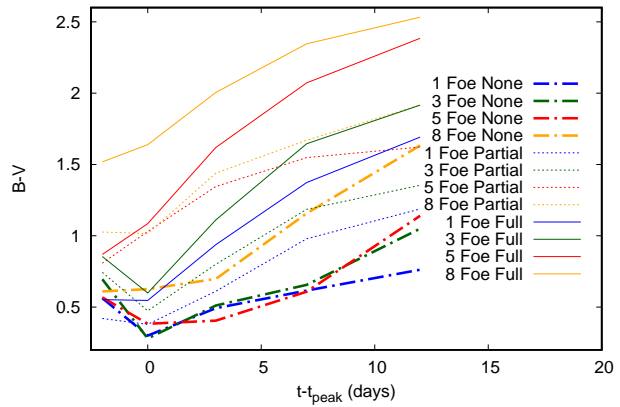}
\caption{The $B-V$ colour evolution of the spectra with all explosion energies and mixing approximations.
    \label{fig:b_min_v}}
\end{figure}

The B-V colour evolution shown in Figure \ref{fig:b_min_v} is mixed with respect to both explosion energy and with the mixing approximations. After t$_{\text{peak}}$, all of the SNe become redder, but at different rates given the mixing approximation. The non-mixed models have a shallower slope, going from an approximate peak B-V of 0.3 for the 1, 3, and 5 foe models at $t_{\text{peak}}$ to a B-V $\sim$ 1.0 - 0.6 at day 12. The 8 foe model has a steeper slope starting from day 3 that matches the day 7 to day 12 slope of the 1, 3, and 5 foe models. The fully mixed models have the steepest slope, with a peak to 12 day B-V change of $\sim$1.0-1.5. This is likely due to the extent of $\nic$ mixing as the centrally located $\nic$ in the non-mixed models heats the ejecta for a longer period of time. The fully mixed models have more $\nic$ further out, which can escape the ejecta without heating the material. 

\begin{figure*}
    \includegraphics[scale=.65,angle=-90]{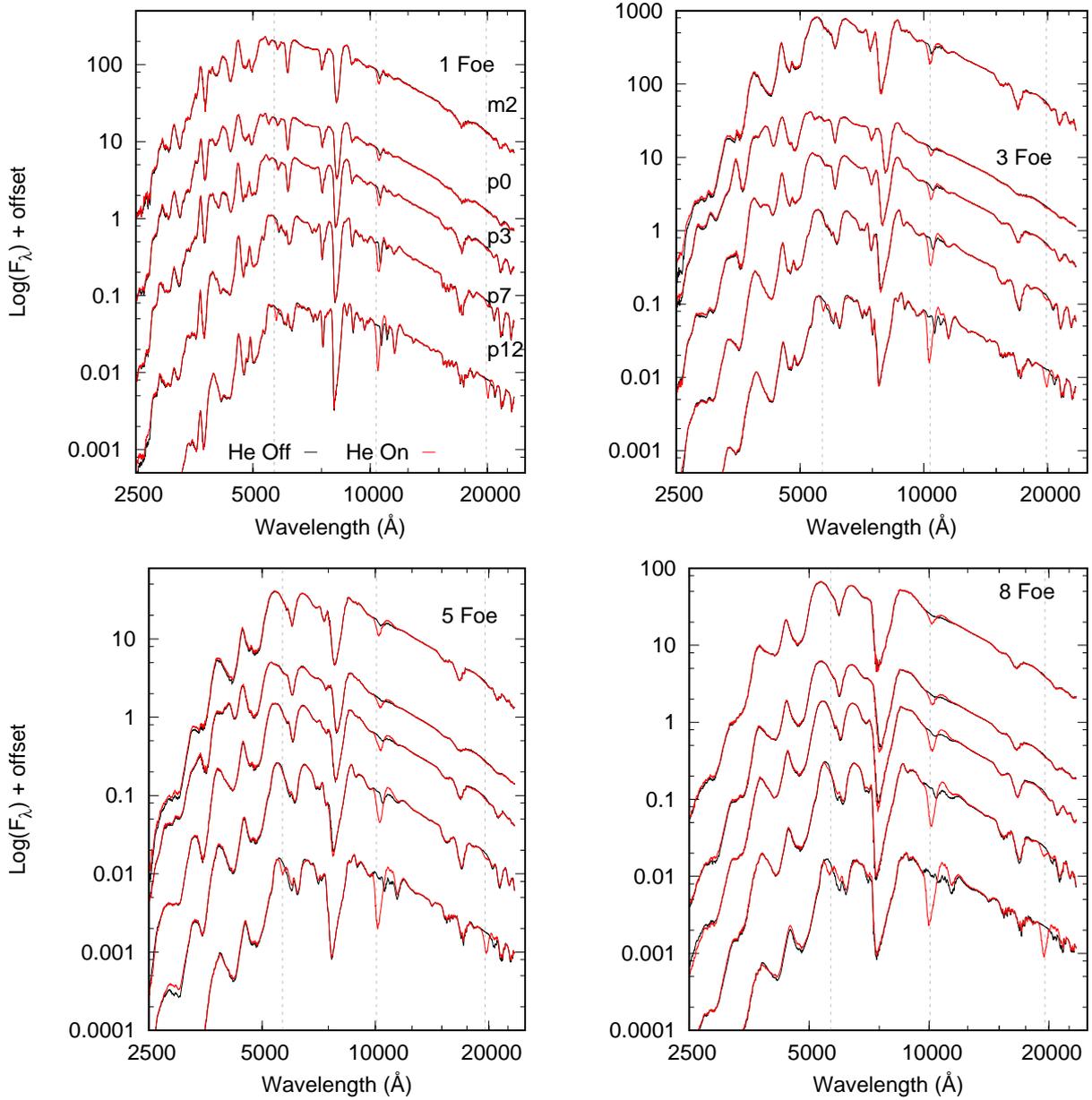}
    \caption{The fully mixed SNe for all four explosion energies. The red line includes the non-thermal effects due to He while the black line shows the same model without the non-thermal effects of He. The dashed grey lines represents three He I lines at rest wavelengths of 5875, 10830, and 20581 $\angstrom$
        \label{fig:log_all_full}}
\end{figure*}

Figure \ref{fig:log_all_full} shows the spectra of only the fully mixed models at all epochs and $E_{\text{k}}$ in order to show how the constant He abundance changes the He features in the spectra. No other mixing approximations are shown as none have helium in the chosen epochs, other than the 8 foe no mixing model. In the previous spectral plots (Figure \ref{fig:spec_all_mix_all_ene}), the wavelength range was plotted from 3000 to 10000 $\angstrom$ to cover the commonly observed range of 4500-9000 $\angstrom$. Two very strong He I features are present at 10830 and 20581 $\angstrom$, however this wavelength range is not often observed by ground based telescopes. At the early epochs, the abundance of He is high enough that the He I line at 10830$\angstrom$ is observable at all explosion energies. This line grows in strength as the ejecta evolves. As the explosion energy increases, the strength of this line is also increased. This is due to the higher energy ejecta placing more He and $\nic$ at higher velocities than in the lower explosion energy models. The 20581$\angstrom$ line is only observable at late epochs in the 1, 3, and 5 foe models and weakly visible in 8 foe models at t$_{\text{peak}}$ + 7 days. The 5875$\angstrom$ line that is commonly used to help classify Type Ib SNe is weak or non existent for most of the epochs, excluding the t$_{\text{peak}}$ + 12 days.

\section{Physical Classification} 
\label{sec:phys_class}

Using the method suggested in \citet{10.1093/mnras/stx980} and discussed in Section \ref{sec:intro}, we categorize our explosion models for all explosion energies and mixing approximations. In this classification system, SNe Ic-BL show strong line blending and are classified as $N=$3-4. These events are often related to GRBs and have high $E_{\text{k}}$. Our spectral classifications are summarized in Table \ref{tab:n_val}. The $N$ values for our spectra show two basic trends. First, as $E_{\text{k}}$ increases, the value of $N$ typically decreases. The spectra for the 8 foe model, for example, show very broad features compared to the narrow lines in the spectra of the 1 foe model. Secondly, for a given $E_{\text{k}}$, the unmixed models have an $N$ value that is one greater than for the fully mixed models while generally sharing a similar $N$ value with the partially mixed models. At the epochs considered the photosphere is still mostly outside the partially mixed region, so that the partially and unmixed models show similar features. After peak the value of $N$ changes somewhat as the photosphere forms at lower velocities and line blending decreases. In particular, the top left plot in Figure \ref{fig:spec_all_mix_all_ene} shows non-blended Fe\,II lines in the spectra of the 1 foe model, changing $N$ from 5(6) to 6(7), where the number in parenthesis is the value of $N$ assuming that Na\,ID is observable. The peak spectra for the no mixing models do not show the Fe\,II multiplet 48 lines but the t$_{peak}$ -2 and +3 both show the lines, so for that mixing approximation, we consider the near peak spectra. 

\begin{table*} 
\begin{tabular}{cccccccccc}
\hline
Energy 	& Mixing? & Fe II   &  Fe II  & Fe II   & Na I & Si II &   O I   & Ca II   & $N$  \\
 & & 4924 $\angstrom$ & 5018 $\angstrom$ & 5198 $\angstrom$ & 5895 $\angstrom$ & 6347 $\angstrom$ & 7774 $\angstrom$ & NIR triplet &  \\
 \hline\hline
1 Foe	& No	  & Yes     & Yes     & Yes	& Yes* &  Yes  & Yes	 & Yes     & 6(7) \\
	    & Partial & Blended & Blended    & Yes	& Yes*  &  Yes  & Yes	 & Yes     & 5(6) \\
        & Full	  & Blended & Blended & Yes	& Yes* &  Yes  & Yes	 & Yes     & 5(6) \\
3 Foe   & No	  & Yes     & Yes     & Yes	& Yes* &  Yes  & Yes	 & Yes     & 6(7) \\
	    & Partial & Yes     & Blended & Blended & Yes* &  Yes  & Yes	 & Yes     & 5(6) \\
        & Full	  & Yes     & Blended & Blended & Yes* &  Yes  & Yes	 & Yes     & 5(6) \\
5 Foe	& No	  & Blended & Blended & Yes	& Yes* &  Yes  & Yes	 & Yes     & 5(6) \\
	    & Partial & Blended & Blended & Yes	& Yes* &  Yes  & Yes	 & Yes     & 5(6) \\
        & Full	  & Blended & Blended & Blended & Yes* &  Yes  & Yes	 & Yes     & 4(5) \\
8 Foe	& No	  & Blended & Blended & Blended & Yes* &  Yes  & Yes	 & Yes     & 4(5) \\
	    & Partial & Blended & Blended & Blended & Yes* &  Yes  & Yes	 & Yes     & 4(5) \\
	    & Full	  & Blended & Blended & Blended & Yes* &  Yes  & Blended & Blended & 3(4) \\	    
\hline
\end{tabular}
\caption{Visibility of the spectral features used in the computation of the $N$ value from \citet{10.1093/mnras/stx980}, and $N$ value for the various models in our set. Lines that are identified at or near peak luminosity, in the t$_{0}$ spectra, and unblended, are labelled with ``yes". If the line appears blended with another line, it is defined as ``blended" and the blend is counted as one feature.  The Na\,ID line is not present in the models because Na is not part of the initial composition of the ejecta. The Na\,ID line is therefore marked as *. This is represented by the number in parenthesis in the N column, where the first number is without Na and the second includes Na. }
\label{tab:n_val}
\end{table*}

\section{Spectral Comparisons} 
\label{sec:fits}

\begin{figure*}
\centering
\includegraphics[scale=1.2]{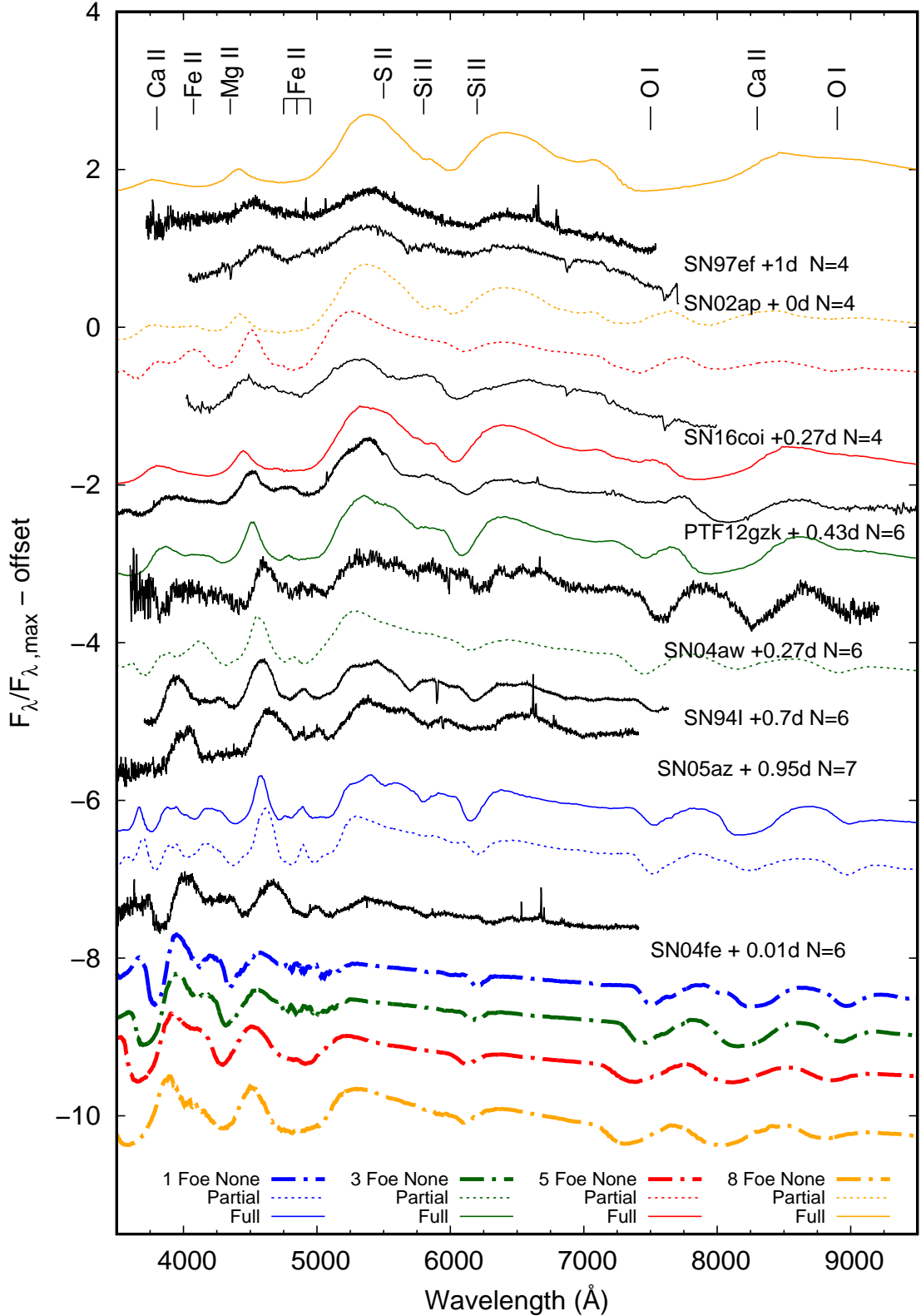}
\caption{Synthetic spectra of all models at bolometric peak compared to a set of observed SNe\,Ic. SNe classified in \citet{10.1093/mnras/stx980} that share $N$ values with those of the models in Table \ref{tab:n_val} are selected. All spectra are arranged in order of colour at peak, with a reddest spectra at the top. SN\,2004aw has been scaled by 1.5x its own maximum for visibility of its features.  As stated in Figure \ref{fig:spec_all_mix_all_ene}, the absorption feature identifiers are approximate. Observational information related to the individual SNe spectra is given in: \citet{2014AJ....147...99M} for SNe 1997ef, 2002ap, 1994I, 2004aw, 2005az, and 2004fe, \citet{2018MNRAS.478.4162P} for SN\,2016coi, and \citet{2016PASA...33...55C} for PTF12gzk. The spectral data were downloaded from WISeREP \citep{2012PASP..124..668Y}. \label{fig:spec_bycolor}} 
\end{figure*}

Having computed synthetic spectra at a number of epochs for all $E_{\text{k}}$ and mixing approximations, we now turn to observational data to check if the results show realistic behaviour. Unlike the previous figures that show the synthetic spectra, we do not plot the logarithmic flux, instead we scale the flux of each SNe, both the synthetic and observed, by the maximum flux value. Figure \ref{fig:spec_bycolor} shows all synthetic spectra at peak luminosity against those of a set of 8 observed SNe.  These 8 SNe are a subset of the SNe that are classified in \citet{10.1093/mnras/stx980}. This subset is chosen because the SNe share $N$ values with those of our synthetic spectra (Table \ref{tab:n_val}). The scaling of the spectra in flux can suppress the impact of luminosity. These factors are an important limitation to keep in mind when trying to make accurate predictions of the ejecta properties based on Figure \ref{fig:spec_bycolor} only. For all of the synthetic spectra shown, we do not include any elements, in particular Na, that were not in the original hydrodynamic model.

The SNe and models in Figure \ref{fig:spec_bycolor} are arranged in order of colour at peak, with redder colours at the top. The $N$ value for the observed SNe is from \citet{10.1093/mnras/stx980}. Because of the sometimes limited wavelength coverage of the spectra, for some SNe it is unclear whether or not the Ca\,II NIR triplet and the O\,I 7774 line are blended, which would affect the value of $N$. In these cases, \citet{10.1093/mnras/stx980} also consider spectra close to, but not exactly at peak. If the two features are present in those spectra, then it is assumed that they would likely appear at peak, with similar profiles, and this information is used to compute $N$.  The unmixed models for all $E_{\text{k}}$  are the bluest of all synthetic spectra, and are bluer than all observed SNe. Thus they are clustered at the bottom despite the $N$ value. The observed $N=4$ SNe, the closest to Type Ic-BL in the sample shown, are the reddest observed SNe and are clustered near the top.  The 8 foe fully and partially mixed models replicate most of the broad features observed in the near-peak spectra of these SNe. Figure \ref{fig:spec_bycolor} shows a slight trend whereby data and models with a smaller $N$ are also redder. At a constant mass, the behaviour of the synthetic spectra reflects the increasing line blanketing in models with larger $E_{\text{k}}$ and more mixing. The observed SNe appear to follow a similar trend, at least with respect to $E_{\text{k}}$, but more data would be helpful in defining the range of properties.

SN\,2016coi shares similar features with the synthetic spectra from the 8 foe fully and partially mixed models, except that the possible Na\,ID line in SN\,2016coi may not be observable in the fully mixed 8 foe model. A mixture of Si and S lines nearby may blend with or swamp the Na\,ID feature, depending on the abundance of Na. The Ca\,II NIR triplet and O\,I 7774 are blended in the fully mixed 8 foe model, but not in the partial and unmixed cases.  \citet{2018MNRAS.478.4162P} obtained an ejecta mass of 4-7\,$\msol$ and $E_{\text{k}}$ $\approx 4.5$-7 foe. The masses of our bare CO cores are at the low end of this range, but the $E_{\text{k}}$ we used, covers the estimated range.  

Of the 8 observed SNe, 5 have $N=6-7$, while 8 of the 12 synthetic spectra have values of $N=6-7$ if we assume that the Na\,ID line would be visible. Spectra with $N=7$ show fully unblended Fe\,II multiplet 48 lines, resulting in three features near 5000\,\AA. The two bluer features, which are closer in velocity space, are blended in $N=6$ spectra. On the observational side, SNe 2004fe and 1994I share a similar set of absorption features from 3900 to 4250\,$\angstrom$. This is reproduced by the partially mixed and unmixed 1 and 3 foe models. SN\,1994I has been modelled to have a low-mass ejecta of $\sim 1\,\msol$ and $E_{\text{k}}$ $\sim 10^{51}$\,erg \citep{Nomoto1994}.  Adopting a combined spectroscopic and photometric approach, \citet{2006MNRAS.369.1939S} found $M_{\text{ej}} \approx 1.1 \msol$ and $E_{\text{k}}$ $\approx 10^{51}$\,erg. The $\nic$ mass was estimated as 0.07\,$\msol$. The requirement that $\nic$ is mixed out to reproduce the light curve was taken as evidence of some asymmetry. No detailed fits are available for SN\,2004fe.  The lack of a GRB connection for these SNe may also suggest that mixing was not as extensive as in GRB/SNe, such that the less mixed model spectra should share more similarities with the data, as they do. 

Figure \ref{fig:spec_bycolor} shows that the scaled synthetic spectra of the 22 $\msol$ progenitor, given the various combinations of $E_{\text{k}}$ and ejecta mixing, can replicate the bulk spectral features of both narrow- and broad-lined SNe. While it may be tempting just to read off values for the properties of the SNe plotted from that figure, this may lead to incorrect results, as we discussed above, because a single $M_{\text{ej}}$ is used and the data are not flux-calibrated. Thus the only information that can reliably be obtained from the comparison is about $E_{\text{k}}$/$M_{\text{ej}}$, which is mostly responsible for the spectra shape \citep{10.1093/mnras/stx992}. While keeping this in mind, we continue the comparison as an exercise.  

SN\,1994I is matched well by the spectrum of the fully mixed 1 foe model and the partially mixed 3 foe model, so we may guess that $E_{\text{k}}$ was 1-3 foe and $E_{\text{k}}$/$M_{\text{ej}}$ $\sim 1$, in accordance with the results of detailed modelling. SN\,2004aw best matches the 3 foe partially and fully mixed models and is in reasonably good agreement with detailed modelling, which also suggests that $M_{\text{ej}}$ was similar to that of the models used here \citep{10.1093/mnras/stx992}. SN\,2002ap \citep{2002ApJ...572L..61M} is placed between the 8 foe partially and fully mixed models, suggesting $E_{\text{k}}$/$M_{\text{ej}}$ $\sim 2$. This is in good agreement with detailed modelling, which however favoured a lower $M_{\text{ej}}$ of $\approx 2.5\,\msol$, and $E_{\text{k}}$ of $\approx 4 \times 10^{51}$\,erg, but the actual ejected mass was smaller than in the models computed here. Finally, SN\,2016coi is also well matched by the partially and fully mixed 5 foe models, but it has not been modelled. 

Given the success of the comparison between synthetic and observed spectra at peak, we can then consider the time evolution of the spectra. Observations of SNe in general may not coincide with the dates of the synthetic spectra. For this reason, the nearest epoch with respect to peak luminosity is used to compare to our synthetic spectra for all observed SNe. 

\begin{figure}
\centering
\includegraphics[scale=.35,angle=-90]{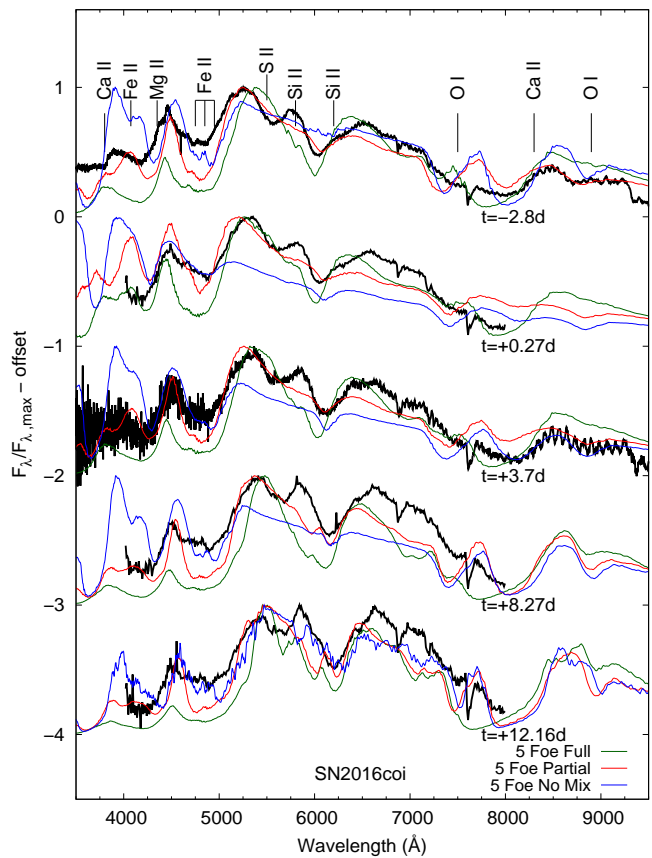}
\caption{The time-series of spectra of the 5 foe model for all three types of mixing compared to the closest available epochs of SN\,2016coi. As stated in Figure \ref{fig:spec_all_mix_all_ene}, the absorption feature identifiers are approximate.
       \label{fig:5foe_16coi}}
\end{figure}

Figure \ref{fig:5foe_16coi} shows the time-series of spectra of SN\,2016coi compared to that of the 5 foe models for the three mixing approximations. None of the spectra match the observed ones perfectly, but the partially and fully mixed models behave better. Only the fully mixed models are able to blend the O\,I 7774 line and the Ca\,II NIR triplet.

\begin{figure}
\centering
\includegraphics[scale=.35,angle=-90]{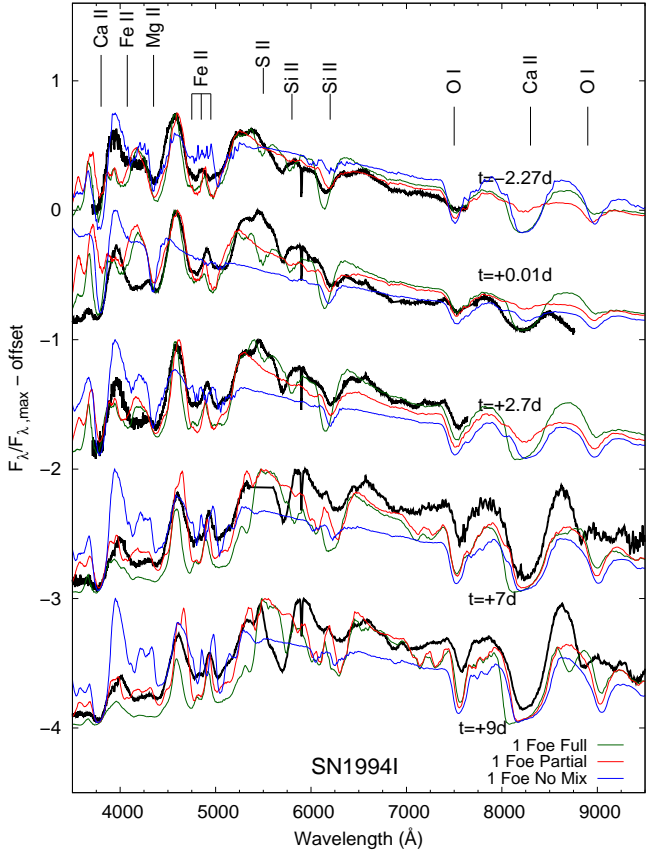}
\caption{The time-series of spectra of the 1 foe model for all three types of mixing compared to the closest available epochs of SN1994I.
\label{fig:1foe_94i}}
\end{figure}

Figure \ref{fig:1foe_94i} shows the 1 foe models for all three mixing approximations and compares them to the time-series of spectra of SN\,1994I. The low velocity ejecta of the 1 foe model replicate the basic features of the observed spectra for all mixing approximations. The Fe\,II line complex is correctly reproduced. The full mixing models best represent these features. This is in agreement with the results of \citet{2006MNRAS.369.1939S}, who do not find large changes in the near-photospheric abundances as a function of time. All mixing approximations also reproduce the O\,I 7774 line and the Ca\,II NIR triplet quite well, except for the possible O\,I line near 9000 $\angstrom$. 

\begin{figure}
\centering
\includegraphics[scale=.35,angle=-90]{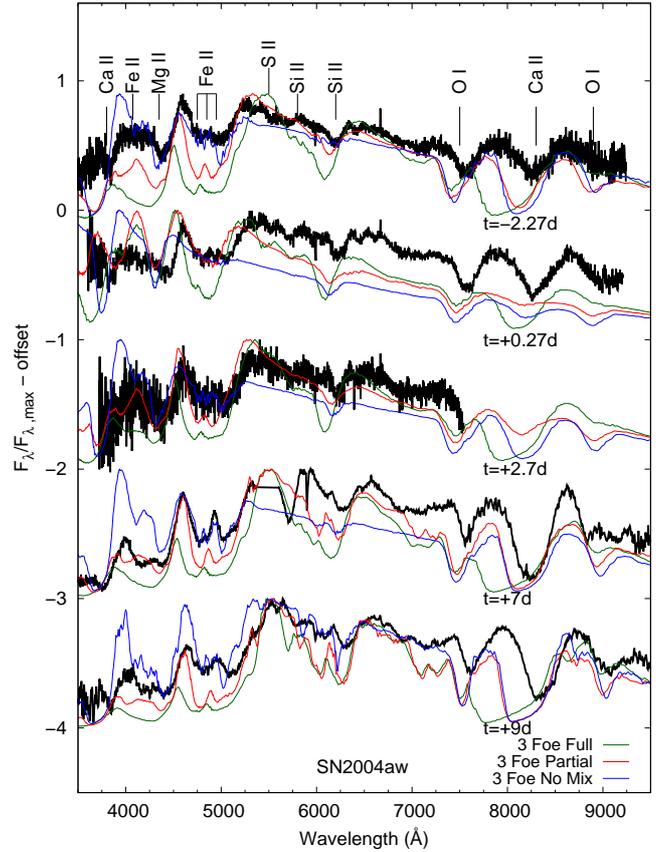}
\caption{The time-series of spectra of the 3 foe model for all three types of mixing compared to the closest available epochs of SN\,2004aw. 
\label{fig:5foe_04aw}} 
\end{figure}

Figure \ref{fig:5foe_04aw} shows the 3 foe models for all three mixing approximations compared to the time-series of spectra of SN\,2004aw. While all models reproduce the observations qualitatively, the partially mixing models appear to be the most accurate, especially in the blue and in the O\,I 7777/Ca\,II NIR triplet region. In the fully mixed models the Si\,II lines near 6000\,$\angstrom$ are significantly too strong throughout the evolution of the ejecta, suggesting this degree of mixing may be too extreme. SN\,2002aw was modelled in detail \citep{10.1093/mnras/stx992}, whose results agree with the values used in these models. 

\begin{figure}
   \centering
   \includegraphics[scale=.35,angle=-90]{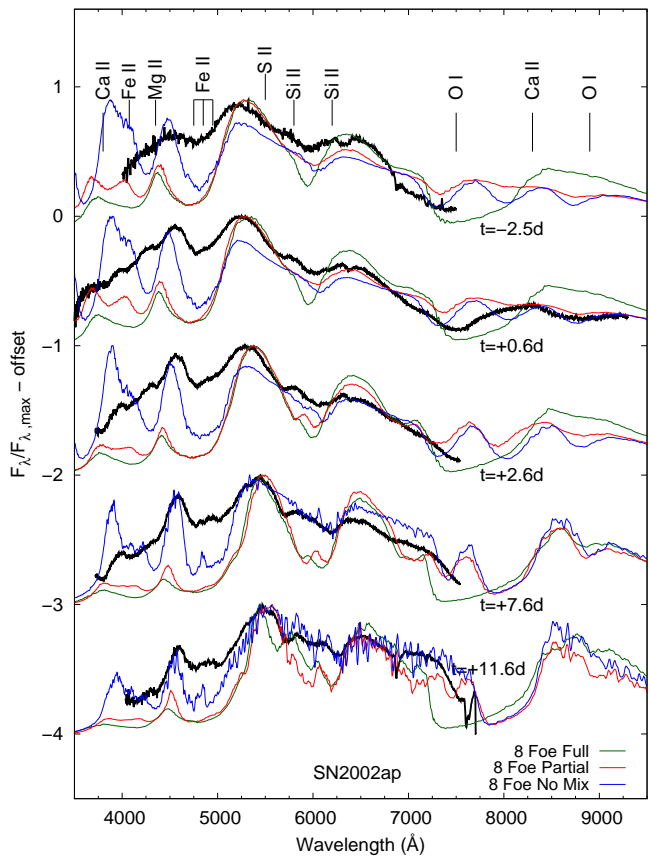}
   \caption{The 8 foe models for all three types of mixing plotted with the closest available epochs of SN2002ap. 
   \label{fig:8foe_02ap}}
\end{figure}

Figure \ref{fig:8foe_02ap} shows the 8 foe models for all three mixing approximations compared to the time-series of spectra of SN\,2002ap. The overall shape of the spectra is best reproduced by the fully and partial mixing models, but the detail match is never perfect, most likely because SN\,2002ap had a smaller ejected mass then the models presented here. Nevertheless, the similarity of the spectral features is striking. All three mixing approximations for the 8 foe model show blended Fe\,II features, but only the fully mixed model blends the Ca\,II NIR triplet and O\,I 7774.  

\begin{figure*}
\centering
\includegraphics[scale=.65,angle=-90]{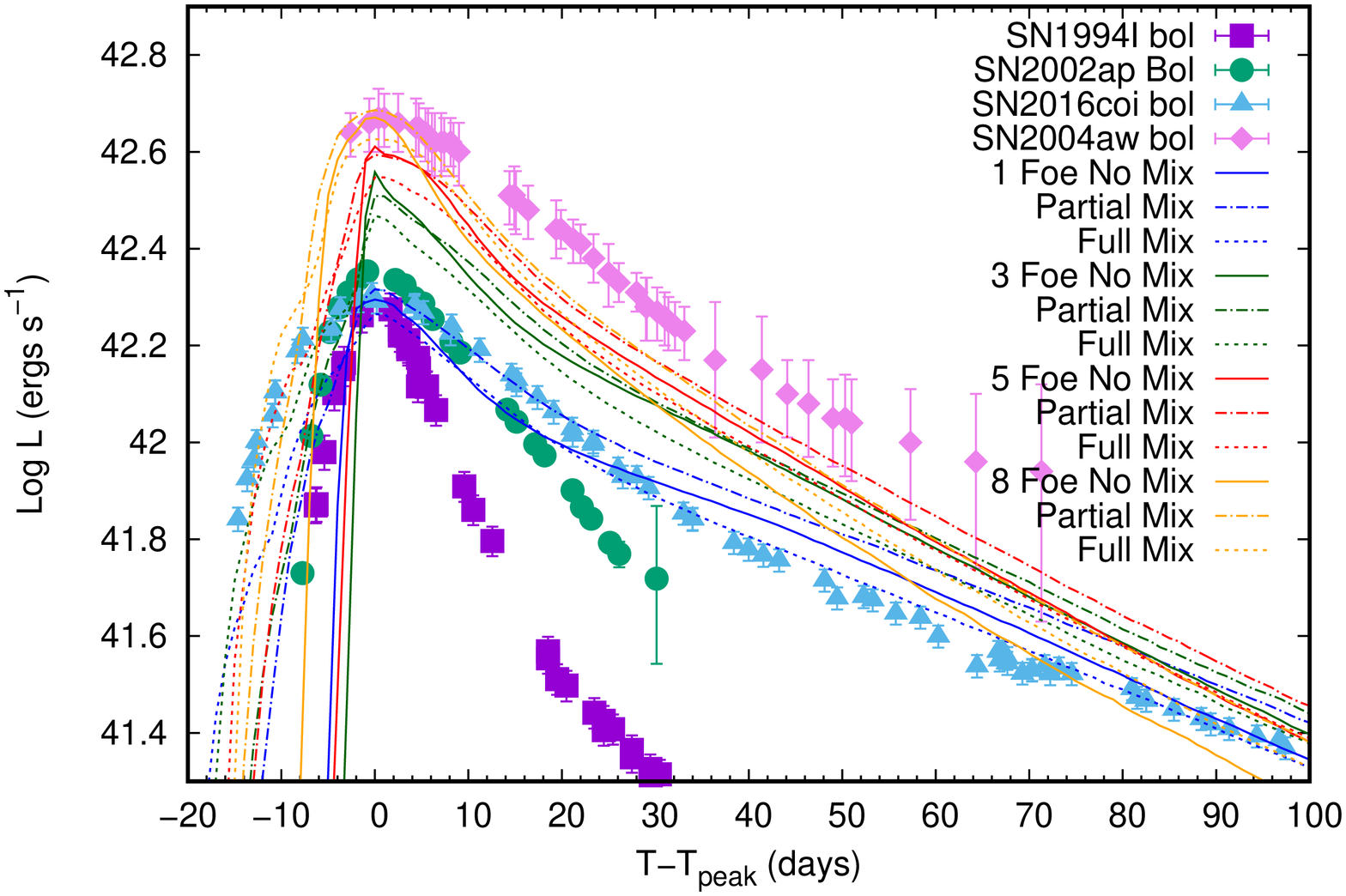}
\caption{The pseudo-bolometric lightcurves of SNe\,1994I, 2004aw, 2002ap, and 2016coi compared to the bolometric light curves of the models for all four explosion energies and all three mixing approximations.  
    \label{fig:all_lc}}
\end{figure*}

Overall, the synthetic and observed spectra do not show a one to one comparison, but this was to be expected. The diversity of Type Ic SNe suggests that they do not share a single unified progenitor and when using models to fit to observed SNe, significant work must be done to identify the most likely progenitor. To expand this, Figure \ref{fig:all_lc} shows the pseudo-bolometric light curves of the four previously mentioned SNe compared to all of our synthetic light curves.  Our initial guess for the $E_{\text{k}}$ of SN\,1994I was in the 1 foe range, however the light curves of the 1 foe models are too broad when compared to SN\,1994I. This suggests that the mass ejected by SN\,1994I was smaller than in our set of models, in accordance with detailed modelling results. The light curve of SN\,2002ap matches best in shape the 3 and 5 foe models, but it did not reach comparably high luminosities. Given the uncertainty of the $\nic$ mass in the models, the peak luminosity could match by choosing a $\nic$ mass that falls within the uncertainty in production.  For SN\,2004aw, the light curve matches best the 5 foe models in shape, although it is somewhat brighter. This $E_{\text{k}}$ is higher than the 3 foe guess suggested by Figure \ref{fig:spec_bycolor}, but both values are not inconsistent with the modelling results. Finally, SN\,2016coi, for which no detailed model exists shows a broad light curve, suggestive of a low $E_{\text{k}}$/$M_{\text{ej}}$ ratio, and is reasonably well matched by the 1 foe models with full mixing, which is at odds with the spectroscopic results above. Accurate modelling of this SN would be desirable. 

\section{Discussion} 
\label{sec:disc}  

The comparisons given in Section \ref{sec:fits} show some similarities in the spectral evolution between the various SNe and the synthetic spectra, but only SNe\,2004aw and (partially) 2016coi also show similarities in their photometric behaviour, as seen in Figure \ref{fig:all_lc}. The fact that synthetic spectra do not perfectly replicate observed SNe is to be expected, as our models were not fine-tuned to fit any specific event. $\nic$ and ejecta mass vary as a function of $E_{\text{k}}$ in our models, but were not free parameters in this work.  The observed light curves suggest some degree of freedom that affects $M_{\text{Ni}}$ was not accounted for in the explosion models as discussed in Section \ref{subsec:abu_prof}. As was noted previously and repeatedly, simply using Arnett's rule to determine bulk properties of the ejecta such as mass and - in particular - $E_{\text{k}}$, may yield a set of values that match the light curve width but are then unable to replicate the spectral features if used in an explosion model.  Looking at the mixing approximations we used, we can also begin to understand how complex the reality of these events. For example, the fully mixed 5 foe explosion replicates some features in SN\,2016coi, which has $N=4$, that the partially and unmixed models cannot match, but no approximation works throughout the observed spectrum at all epochs. The fully mixed models are uniform in composition as shown in Figure \ref{fig:abu_full}, and the divergence from the observed SNe suggests that in reality mixing is not as uniform. SN\,2004aw ($N=6$) was a moderately high energy SNe but with no co-incident GRB. In this case, the fully mixed 5 foe explosion model has features that are too broad when compared to SN2004aw's spectra. The partially mixed model has better success. This is also backed by the fitted abundance structure in \citet{10.1093/mnras/stx992}, which shows fairly strong $\nic$ and Fe group mixing outside the inner regions. This type of mixing is seen in Figure \ref{fig:abu_mixing}. Unlike our partial mixing approximation, IME are mixed further out into the ejecta. SN\,1994I, which is also well studied, also required some mixing of $\nic$ and Fe group elements outward in order to replicate the light curve \citep{2006MNRAS.369.1939S}. 

The comparison between our model set and observed SN\,Ic data gives more evidence that analyzing observed SNe requires a combination of spectroscopic and photometric methods to get a solid estimate of the ejecta parameters. The methods in this work are limited as SNe\,Ic come with a range of $M_{\text{ej}}$, $E_{\text{k}}$, and $M_{\text{Ni}}$. Additionally, when using scaled spectra one cannot account for luminosity differences between the observed and synthetic spectra. The spectral similarities between the observed and synthetic spectra do give suggestions about the $E_{\text{k}}$/$M_{\text{ej}}$ ratio. For the same $\nic$ mass, multiple similar light curves can be generated by changing $E_{\text{k}}$ and $M_{\text{ej}}$ while keeping $E_{\text{k}}$/$M_{\text{ej}}$ constant. 

Some of the observed light curves in Figure \ref{fig:all_lc} show a much broader rising phase compared to the unmixed models calculated lightcurve. The mixed models, however, reproduce this early rise much better. This shows that the width of the light curve alone cannot simply be used to determine $M_{\text{ej}}$ and $E_{\text{k}}$. In this study, $M_{\text{ej}}$ was nearly constant for 4 explosion energies and associated $\nic$ masses. The only ``free" parameters we had is how we altered the interior density structure and mixed the $\nic$ and other elements. Modifying the distribution of $\nic$ leads to significant changes in the width of the light curve, while $M_{\text{ej}}$ and $E_{\text{k}}$ are constant. In mixed models, the presence of $\nic$ at higher velocities causes an earlier rise of the luminosity, and thus broadens the light curve. When comparing the mixed to the unmixed models, the broadening in the rise is significant for the same $E_{\text{k}}$/$M_{\text{ej}}$. This shows that using Arnett's approximation (which assumes all $\nic$ to be centrally located as a point source) without taking into account other factors that influence photon diffusion time, and hence affect the light curve shape,  can lead to incorrect estimates of the SN parameters as thoroughly discussed in \citet{10.1093/mnras/stx992}.

\citet{10.1093/mnras/sty3399} give rise times for SNe\,Ib/c in the range from $8.6\pm _{2.0}^{3.8}$ to $10.4\pm _{1.7}^{2.8}$ days. SNe\,Ic have shorter rise times than SNe\,Ib. SESNe with steeply rising light curves, such as those of the 1, 3, and 5 foe models shown in Figure \ref{fig:lc_nomix} may be absent from observational studies such a rapid rise is difficult to observe. Whether totally unmixed explosions with a structure similar to that of our models do indeed exist is an interesting question. Section \ref{sec:fits} shows that both fully and partially mixed models can replicate features much better across multiple epochs for various observed SNe. 

\citet{2019ApJ...872..174Y} discusses the affect of $\nic$ mixing in the very early phases of the explosion. Their code and work considers phases with respect to peak of -10 to -25 days and include the contribution of breakout emission. Our code does not treat this mechanism or consider phases this early. Both light curves and spectral formation are highly dependent on the explosion models used and the models in our work are quite different from the ones in \citet{2019ApJ...872..174Y}. As such we cannot make strong comparisons to their results but we will consider a few aspects. They do show that as the mixing of $\nic$ increases, the rise time increases, which matches our work. The B-V values in \citet{2019ApJ...872..174Y} show that the least mixed models have a lower B-V at peak and the fully have the highest B-V at peak if we only consider the time range of only -2 to 12 days from peak, similar to the time range in our work. This matches our results in Figure \ref{fig:b_min_v}.

\begin{figure}
\centering
\includegraphics[scale=.33,angle=-90]{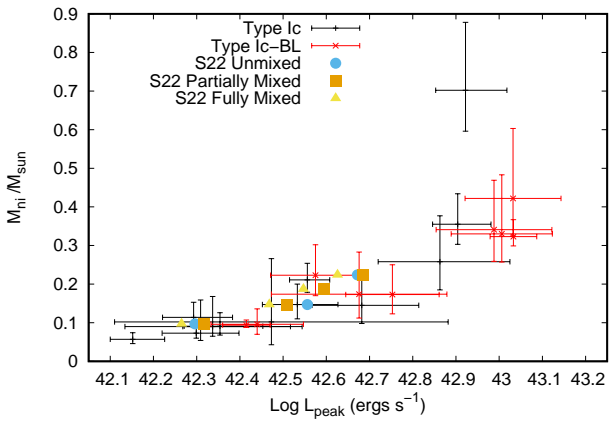}
\caption{The peak luminosity calculated from a set of fully bolometric light curves taken from \citet{doi:10.1093/mnras/stw299} compared to the measured $\nic$ mass using Arnett's rule. The $\nic$ mass in some SNe is a lower limit because the host extinction correction is not available. 
    \label{fig:lp_v_ni}} 
\end{figure} 

Figure \ref{fig:lp_v_ni} shows the peak bolometric luminosity and the derived $M_{\text{Ni}}$ for a set of SNe\,Ic from \citet{doi:10.1093/mnras/stw299}, along with $M_{\text{Ni}}$ of our s22 models and their computed peak luminosity. The slight changes in peak luminosity caused by the different degrees of mixing in the models are shown. The fully mixed models have the lowest peak luminosity because the early diffusion of photons reduces the radiative energy available at peak. The peak luminosities of the partially mixed and unmixed models are similar. These two mixing methods keep the bulk of $\nic$ in the interior region, although with different densities (Figure \ref{fig:dens}). The $\nic$ yields are somewhat uncertain in our explosion models as discussed in Section \ref{subsec:abu_prof}, but there is likely a correlation between $E_{\text{k}}$ and $M_{\text{Ni}}$, which reproduced observational results. Peak luminosities mostly reflect $M_{\text{Ni}}$, even though $M_{\text{ej}}$ changes somewhat for different $E_{\text{k}}$.

Because of the approximate treatment of nucleosynthesis in the supernova simulations, using a 13-species $\alpha$-network and a 15-species solver for nuclear statistical equilibrium, the exact amount of $^{56}$Ni ejected in the explosions is unclear. A considerable fraction of tracer material (X$_{56}$) and of the $\alpha$-particles of the neutrino-heated ejecta could also be $^{56}$Ni, see the detailed discussion in Section 3.3 of \citet{2019arXiv191001641E}.

Mixing $\nic$ and iron to higher velocities produces line blanketing effects in the bluer portion of the spectra, particularly the near-UV. This is shown in the extreme case for the higher $E_{\text{k}}$ fully mixed models, where the region between 2000 and 5000\,$\angstrom$ is mostly blanketed.  This could suggest that the extent of mixing may be derivable by this region of the spectra, as it has been for SNe Ia \citep{2014MNRAS.439.1959M}, but this region is rarely covered observationally.

With four explosion energies and a single mass model, we tried to cover the likely range of observed $E_{\text{k}}$/$M_{\text{ej}}$ (Table \ref{tab:model_params}). The $N$ values shown in Table \ref{tab:n_val} cluster around 5-6 with the 5 and 8 foe models having lower values. As discussed in \citet{10.1093/mnras/stx980}, the number of observed SNe for which $E_{\text{k}}$/$M_{\text{ej}}$ is well defined (i.e. it was obtained from detailed modelling) is low, but still a trend appears to exist whereby $N$ increases as $E_{\text{k}}$/$M_{\text{ej}}$ decreases, as simple considerations also suggest. This is reflected in our models. Mixing has a tendency to favour line blending (the Fe\,II lines, Ca\,II IR triplet and O\,I 7774) and thus to reduce the value of $N$.

Figure \ref{fig:log_all_full} shows that fully mixing the ejecta, including the He remaining after $\alpha$-rich freeze-out, can produce NIR He lines of varying strengths with respect to $E_{\text{k}}$, but may not produce a strong enough optical He I line for early or easy classifications to a Type Ib. Observing the 10830$\angstrom$ He I line early enough may be necessary for clear differentiation between Type Ib and Ic SNe but this is complicated by the Mg II line in this region. This requires detailed modelling and wider wavelength observations at early and late times if the optical He I lines are weak or hidden. 

The lack of $E_{\text{k}}$/$M_{\text{ej}}$ ratios inferred using hydrodynamic models coupled with radiative transfer simulations makes drawing comparisons from that ratio to quantities such as $M_{\text{Ni}}$, $M_{\text{ej}}$, and $E_{\text{k}}$ challenging. Unlike analytic methods, hydrodynamic models coupled with radiative transfer simulations can be time consuming due to the uncertain nature and interconnected behaviours that occur when changing variables. This is where a parameter study can fit in as an extension to available data to determine how variables can affect observable properties \citep{Young_2004}.  As shown in this work, a single mass with a single $E_{\text{k}}$ but variable mixing can produce SN spectra with a different $N$ value. Using multiple models, with different $M_{\text{ej}}$, variable $E_{\text{k}}$, and a range of mixing approximations can generate a spectral and photometric library that may constrain some parameters. A bulk study of these initial variables combined with physically reasonable and tunable parameters can be used to generate a wide set of possible progenitors that then can be compared to observed photometry and spectra. 

Turbulent mixing can be approximated in 1-D, but large scale asymmetries or 3D mixing processes may be underestimated. In this work, a density spike in the original explosion models led to unrealistic light curves and spectra for those models. Therefore, the spike was smoothed out when mixing was applied. However, altering the model in such a way may nullify the predictive power of an evolutionary model. The time and computational cost required to model turbulence accurately in 2- or 3-D simulations limits the ease with which explosion models can be generated to match observed SNe. Implementing the Rayleigh-Taylor instability formulation in 1-D \citep{0004-637X-821-2-76} may help produce an evolutionary model that can match spectroscopic and photometric SN data.

\section{Conclusions} 
\label{sec:conc}

In this work, we exploded a 22 $\msol$ progenitor stripped to a bare CO core with $E_\text{k}$ of 1, 3, 5, and 8 foes using the 1-D explosion code \textsc{Prometheus-HOTB}. For each explosion model we alternatively do not mix, partially mix, or completely mix the composition of the ejecta and the innermost density. For each combination of $E_{\text{k}}$ and mixing, we generated a light curve and a time-series of spectra. Each model was then classified using the methods from \citet{10.1093/mnras/stx980} and compared to SNe with similar classifications. 

We showed that the extent of the mixing of Fe group elements and $\nic$ in the ejecta of a SESNe alters both the light curve and the spectra. Hydrodynamic explosion models, with no modification, are characterized by a positive density gradient at the edge of the $\nic$-forming region that can prevent the escape of photons from the deposition of $\nic$/$^{\text{56}}$Co decay products. Smoothing out the structure of this interior density region and mixing $\nic$ outwards into the ejecta produces much broader light curves.

We also show that the combination of mixing the ejecta and increasing $E_{\text{k}}$ leads to SNe with broader lines (Figure \ref{fig:spec_bycolor}). Broad-lined SNe may be the combined result of high $E_{\text{k}}$, significant mixing, and ejecta mass. SNe\,Ic-BL are the only ones that correlate with long GRB, which are high energy events that often include a jet \citep{Woosley1997,10.1093/mnras/stz1588}. Jet-driven SNe material likely results in more mixing than found in SNe with smaller $E_{\text{k}}$. 

Future work would expand this analysis towards Type Ib/IIb SNe. The formation of He lines are sensitive to the mass and mixing of the $\nic$ in the ejecta \citep{1991ApJ...383..308L,2012MNRAS.422...70H}. Depending on the mass of He in the outer ejecta, the mixing of $\nic$ may determine whether or not He is visible during the spectral evolution. This is a partial continuation of work done by \citet{2012MNRAS.422...70H} in which the mass of He or H present in the outer layers of the star was altered to determine the lower limits of possible H/He masses given some constraints. Continuing this work on SNe\,Ic, additional grid points with respect to mass will be considered using progenitors with smaller and larger masses. This would expand the parameter space explored, covering most of the measured ejecta masses using for the same grid of explosion energies and mixing approximations.

\section*{ACKNOWLEDGMENTS}
The authors would like to thank Simon Prentice for useful discussion and observational data. At Garching, this project was supported by the European Research Council through grant ERC-AdG No.341157-COCO2CASA, and by the Deutsche Forschungsgemeinschaft through Sonderforschungbereich SFB 1258 `Neutrinos and Dark Matter in Astro- and Particle Physics' (NDM) and the Excellence Cluster Universe 'ORIGINS: From the Origin of the Universe to the First Building Blocks of Life' (EXC 2094; \url{https://www.origins-cluster.de})
\bibliographystyle{mnras}
\bibliography{sources_file_pap1}







\bsp	
\label{lastpage}
\end{document}